\documentclass[twocolumn]{aastex61}
\usepackage{hyperref}
\usepackage{nth}
\defcitealias{chempy}{R17}
\defcitealias{2018ApJ...861...40P}{P18}

\newcommand{\resub}[1]{#1}
\let\vec\mathbf

\usepackage{tikz}
\usepackage[caption=false]{subfig}
\usetikzlibrary{bayesnet}

\submitjournal{ApJ}

\shorttitle{Inferring Galactic Parameters from Chemical Abundances}
\shortauthors{O.\,H.\,E.\,Philcox \& J.\,Rybizki}

\begin{document}

\title{Inferring Galactic Parameters from Chemical Abundances: A Multi-Star Approach}

\correspondingauthor{Oliver H.\,E. Philcox}
\email{ohep2@alumni.cam.ac.uk}

\author[0000-0002-3033-9932]{Oliver H.\,E. Philcox}
\affil{Department of Astrophysical Sciences, Princeton University, Princeton, NJ 08544, USA}
\affil{Center for Astrophysics | Harvard \& Smithsonian, 60 Garden St., Cambridge, MA 02138, USA}

\author[0000-0002-0993-6089]{Jan Rybizki}
\affil{Max Planck Institute for Astronomy, K\"onigstuhl 17, 69117 Heidelberg, Germany}



\begin{abstract}
Constraining parameters such as the initial mass function high-mass slope and the frequency of type Ia supernovae is of critical importance in the ongoing quest to understand galactic physics and create realistic hydrodynamical simulations. In this paper, we demonstrate a method to precisely determine these using individual chemical abundances from a large set of stars, coupled with some estimate of their ages. Inference is performed via the simple chemical evolution model \textit{Chempy} in a Bayesian framework, marginalizing over each star's specific interstellar medium parameters, including an element-specific `model error' parameter to account for inadequacies in our model. Hamiltonian Monte Carlo (HMC) methods are used to sample the posterior function, made possible by replacing \textit{Chempy} with a trained neural network at negligible error. The approach is tested using data from both \textit{Chempy} and the IllustrisTNG simulation, showing sub-percent agreement between inferred and true parameters using data from up to 1600 individual stellar abundances. For IllustrisTNG, strongest constraints are obtained from metal ratios, competitive with those from other methods including star counts. Analysis using a different set of nucleosynthetic yields shows that incorrectly assumed yield models can give non-negligible bias in the derived parameters; this is reduced by our model errors, which further show how well the yield tables match data. We also find a significant bias from analyzing only a small set of stars, as is often done in current analyses. The method can be easily applied to observational data, giving tight bounds on key galactic parameters from chemical abundances alone.
\end{abstract}

\keywords{astrochemistry --- ISM: abundances, evolution --- Galaxy: fundamental parameters --- methods: statistical --- stars: abundances}

\section{Introduction}
The construction of steadily more accurate large-scale galactic and cosmological simulations is an ongoing effort in the astronomical community \citep[e.g.][]{2012A&A...547A..63F,2017MNRAS.467..179G,2018MNRAS.475..648P}, yet all of these rest upon potentially unjustified assumptions about the values of galactic parameters which control a number of effects, including the birth and death rates for various types of stars. Two crucial unknowns are the shape of the initial mass function (IMF), setting the mass distribution of stars born from the interstellar medium (ISM), and the rate of Type Ia supernovae (SN\,Ia) explosions.

Despite a wealth of work on the subject, the constraints on these parameters remain weak, although it is clear that their values play an important part in determining chemical evolution tracks \citep{2005A&A...430..491R,2015MNRAS.449.1327V,2015MNRAS.451.3693M}. For example, a large variety of high-mass IMF slopes have been posited \citep[Tab.\,7]{2016ApJ...824...82C}, with a steeper-than-canonical gradient being suggested by a range of studies using varied data-sets including M31 star counts \citep{2015ApJ...806..198W}, galactic disk structure \citep{2015MNRAS.447.3880R} and analysis of thin-disk stars \citep{2014ApJ...796...75C}. In addition, the IMF slope may itself be a function of metallicity, introducing further complexity \citep[e.g.][]{2019MNRAS.482..118G,2019A&A...626A.124M}. There is also contention regarding the choice of SN\,Ia delay-time-distribution and normalization \citep{2010ApJ...722.1879M,2012MNRAS.426.3282M,2015ApJ...810..137J}, which plays a crucial role in the enrichment of the ISM.

Given the growing wealth of stellar observational abundance data \citep[e.g. from APOGEE;][]{2016AN....337..863M}, this would seem to be a key data-set with which to constrain galactic parameters, and previous work has contributed to this, utilizing either chemical abundances from a small set of stars (\citealt{chempy}, hereafter \citetalias{chempy}) or entire chemical evolution tracks \citep{2015MNRAS.451.3693M,2018arXiv180208432R}, although only through use of binned statistics. Many of these analyses are unable to implement a fully Bayesian approach, which has the advantage of giving numerical constraints with the ability to marginalize out nuisance parameters. Thanks to the relatively tight bounds that can now be placed on stellar ages \citep{2016ApJ...823..114N,2016MNRAS.456.3655M,2016ApJ...817...40F}, we may begin to explore the huge expanses of data provided by the individual chemical abundances of a large set of stars,  which can be used to place strong constraints on galactic parameters.

The principal goal of this work is to demonstrate how we may use modern statistical techniques and machine learning in tandem with a simple galactic chemical evolution (GCE) model in a Bayesian framework to infer global galactic parameters from a set of stars. We will focus on two key parameters; the high-mass slope of the \citet[Tab.\,1]{2003PASP..115..763C} IMF and the rate of SN\,Ia explosions per unit mass, both of which we assume to be constant across the galaxy. Our primary framework will be based around the \textit{Chempy} model, a simple GCE parametrization that is able to predict stellar chemical abundances given a number of galactic parameters. Previous work with \textit{Chempy} (\citetalias{chempy}; \citealt{2018MNRAS.477.2326F}; \citealt{2018ApJ...861...40P}, hereafter \citetalias{2018ApJ...861...40P}) has concentrated on its application to proto-solar abundances; here we aim to extend this by using multiple stellar data-points. The larger volume of data should be able to give tighter statistical constraints on those parameters that are held fixed across the galaxy, but complexity is added since we must allow each star to carry its own set of local ISM parameters.

Our inference will make use of the modern statistical technique Hamiltonian Monte Carlo (HMC; \citealt{2012arXiv1206.1901N}) sampling, made possible by the replacement of the \textit{Chempy} function with a trained neural network following \citetalias{2018ApJ...861...40P}. For high-dimensional posterior functions, HMC gives much faster sampling than conventional Markov Chain Monte Carlo (MCMC) methods, with the neural network allowing for analytic differentiability. We will test our analysis using mock observations drawn firstly from \textit{Chempy} then from large-scale hydrodynamical simulations to ensure that we recover the correct parameters even for models with a completely different treatment of ISM physics. The methods could naturally be extended to any fast and flexible GCE model, not just \textit{Chempy}. The code used in this paper builds upon the \textit{ChempyScoring} module \citepalias{2018ApJ...861...40P} and has been made publicly available as a new package, \textit{ChempyMulti} \resub{\citep{oliver_philcox_jan_rybizki_2019}},\footnote{\href{https://github.com/oliverphilcox/ChempyMulti}{github.com/oliverphilcox/ChempyMulti}} including a comprehensive tutorial covering both the \textit{Chempy} model and HMC inference.


We begin by describing the GCE models in Sec.\,\ref{sec: GCE_models}, before considering how to use machine learning to optimize the latter in Sec.\,\ref{sec: neural_nets}. Secs.\,\ref{sec: stat_model}\,\&\,\ref{sec: sampling} discuss the Bayesian statistics and our methods to sample from them, before we present the results for two sets of mock data in Sec.\,\ref{sec: results}. We conclude with a summary in Sec.\,\ref{sec: conclusion}. In the appendix, we give technical details of the neural network, a general overview of HMC sampling and representative sampling plots in sections \ref{appen: neural_network}-\ref{appen: full_corner_plot} respectively.

\section{Galactic Chemical Evolution Models}\label{sec: GCE_models}

In order to infer galactic chemical evolution (GCE) parameters we need a simple physical model that takes these as inputs and can be inserted into a Bayesian framework. In addition, if we are interested in testing the validity of our approach, we require a high-resolution simulation which (a) has outputs which may be used in place of observational data (in the form of stellar ages and proto-stellar abundances) and (b) has well-defined values of the global parameters that we can compare to those inferred. Galactic-scale hydrodynamical simulations can be effectively used in this context. We thus need two independent GCE models in our analysis, of high- and low-complexity respectively.

\subsection{IllustrisTNG}\label{subsec: TNG_intro}
In this paper, we use mock observational data derived from the IllustrisTNG (hereafter TNG) magnetohydrodynamical simulations \citep{2018MNRAS.475..624N,2018MNRAS.475..648P,2018MNRAS.480.5113M,2018MNRAS.477.1206N,2018MNRAS.475..676S,2019ComAC...6....2N}.\footnote{\href{http://www.tng-project.org/}{www.tng-project.org/}} These are a successor to the Illustris simulations \citep{2014Natur.509..177V,2015A&C....13...12N}, using an updated physical and chemical model, including new galactic physics and an improved set of nucleosynthetic yields. Here, we are principally interested in the TNG100-1 simulation (of dimension $L\sim110\,\text{Mpc}^3$) which provides the highest resolution publicly available data, at a baryonic mass resolution of $1.4\times10^6\,\mathrm{M}_\odot$ \citep{2019ComAC...6....2N}. Importantly, both the high-mass slope of the \citet[Tab.\,1]{2003PASP..115..763C} IMF and the SN\,Ia normalization \resub{(equal to the number of SN\,Ia formed in $13.8\,\mathrm{Gyr}$ per unit mass)} are fixed parameters in TNG, with values $\alpha_\mathrm{IMF}=-2.3$ and $N_\mathrm{Ia}=1.3\times10^{-3}\,\mathrm{M}_\odot^{-1}$ respectively \citep{2018MNRAS.473.4077P}. 

The simulation consists of a vast amount of galaxies (clustered in dark matter halos), each of which hosts a large number of sub-particles, which can be considered as different stellar environments, subject to some set of latent parameters describing the inter-stellar medium (ISM) therein. For each sub-particle, TNG records the typical birth-time of a star in this location, as well as its initial abundances, thus this provides an excellent set of mock stellar abundance data. This is similar to that found in a typical observational data-set such as the APOGEE catalog \citep{2016AN....337..863M}, but no post-birth abundance corrections are required. This data, coupled with the fixed galactic parameters, allows us to test the validity of our full analysis pipeline including the approximations made by our simple GCE model used for Bayesian inference. 

\subsection{\textit{Chempy}}\label{subsec: chempy_intro}
\textit{Chempy} \citepalias{chempy} is a simple one-zone GCE model that computes the chemical evolution of a region of the ISM throughout cosmic time.  Through use of published nucleosynthetic yield tables for three key processes (SN\,Ia and SN\,II explosions and AGB stellar feedback) and a small number of parameters controlling simple stellar populations (SSPs) and ISM physics, the model predicts ISM chemical element abundances at time $T$, which can be \resub{matched to} proto-stellar abundances for a star born at the same time $T$ \resub{that act as a proxy for the ISM abundances}. Despite its simplicity, the model has been shown to work well in a variety of contexts \citep[e.g.][]{2018MNRAS.477.2326F}, especially due to its speed. As discussed below, this speed is greatly boosted by use of machine learning, first demonstrated in \citetalias{2018ApJ...861...40P}. 

Here, we allow six \textit{Chempy} parameters to vary freely, as shown in Tab.\,\ref{tab:priors}. These may be split into three groups:
\begin{enumerate}
    \item $\vec\Lambda$: \textbf{Global Galactic Parameters}. These describe SSP physics, and comprise the high-mass \citet{2003PASP..115..763C} IMF slope, $\alpha_\mathrm{IMF}$, and (logarithmic) Type Ia supernovae normalization, $\log_{10}(N_\mathrm{Ia})$. We assume these to be constant across both the variety of ISM environments found in a typical galaxy and cosmic time, thus are treated as star-independent in this analysis. (Whilst $\log_{10}(N_\mathrm{Ia})$ is constant with respect to time by definition, it being simply a normalization constant, there is some evidence for $\alpha_\mathrm{IMF}$ varying as a function of time or metallicity \citep{2014ApJ...796...75C,2016MNRAS.462.2832C,2019MNRAS.482..118G,2019A&A...626A.124M}, though this is not included in the TNG model.) We adopt the same broad priors as \citetalias{2018ApJ...861...40P} for these variables (as stated in Tab.\ref{tab:priors}), noting that these fully encompass the values chosen by the TNG simulation.  
    \item $\{\vec\Theta_i\}$: \textbf{Local Galactic Parameters}. These describe the local physics of the ISM and are hence specific to each stellar environment, indexed by $i$. As defined in \citetalias{chempy}, these include the star-formation efficiency (SFE) parameter, $\log_{10}(\text{SFE})$, $\log_{10}(\mathrm{SFR}_\mathrm{peak})$, which controls the peak of the star formation rate (SFR), and the outflow feedback fraction, $\mathrm{x}_\mathrm{out}$ \resub{(controlling the fraction of stellar outflow that is fed to the simulation gas reservoir; the remainder enriches the local ISM)}. We adopt broad priors for all parameters and, as in \citetalias{2018ApJ...861...40P}, do not allow the SN\,Ia delay-time distribution to vary freely, fixing it to the TNG form.
    \item $\{T_i\}$: \textbf{Stellar Birth-Times}. This is the time in Gyr at which a given star is formed from the ISM, and we assume that its proto-stellar abundances match the local ISM abundances at $T_i$. Unlike in previous \textit{Chempy} analyses, this is required to be a free parameter (since it is rarely known to high precision), and we adopt individual priors from mock observational data for each star. (For real data-sets, we can use the computed age estimates to define this, e.g. \citealt{2016ApJ...823..114N}). The \textit{Chempy} code has been adapted to take this as an input, allowing the simulation to stop and return abundances at $T_i$.
\end{enumerate}

\resub{The separability of local (ISM) parameters and global (SSP) parameters is motivated by recent observational evidence: \citet{2019arXiv190710606N} find that the elemental abundances of red clump stars belonging to the thin disk can be predicted almost perfectly from their age and [Fe/H] abundance. This implies that the key chemical evolution parameters affecting the elemental abundances (SSP parameters and yield tables) are held fixed, whilst ISM parameters vary smoothly over the thin disk (which offsets the metallicity for different galactocentric radii). Similarly \cite{2019ApJ...874..102W} find that ISM parameter variations are deprojected in the [X/Mg] vs [Mg/H] plane (their Fig.\,17) and that abundance tracks in that space are independent of the stellar sample's spatial position within the Galaxy (their Fig.\,3).}

To avoid unrealistic star formation histories (that are very `bursty' for early stars), we additionally require that the SFR (parametrized by a $\Gamma$ distribution with shape parameter $a=2$) at the maximum possible stellar birth-time ($13.8$\,Gyr) should be at least 5\% of the mean SFR, ensuring that there is still a reasonable chance of forming a star at this time-step. This corresponds to the constraint $\log_{10}\left(\mathrm{SFR}_\mathrm{peak}\right)>0.294$.\footnote{In analyses using, for example, a set of old stars, this restriction is not appropriate, since it forces there to still be a non-negligible SFR today. In these cases, the condition should be relaxed.} For this reason, a truncated Normal prior will be used for the SFR parameter. Furthermore, we constrain $T_i$ to the interval $[1,13.8]$\,Gyr (assuming a universe age of 13.8\,Gyr as in the TNG cosmology), ignoring any stars formed before $1$\,Gyr, which is justified as these are expected to be rare.

\begin{tiny}
\begin{table*}
\begin{minipage}{\textwidth}
\begin{center}
\caption{Free \textit{Chempy} parameters for each star, with their prior values and Gaussian widths. Prior parameters for stellar birth-times are set for each star individually, based on realistic age estimates, assuming 20\% errors.}
\begin{tabular}{ rl|ccc }
Parameter & Description & $\overline{\theta}_\mathrm{prior}\pm\sigma_\mathrm{prior}$ & Limits & Approximated prior based upon: \\
 \hline
  \multicolumn{5}{c}{$\vec{\Lambda}$: \textit{Global stellar (SSP) parameters}}\\
\hline
  $\alpha_\mathrm{IMF}$ & High-mass slope of the \citet{2003PASP..115..763C} IMF & $-2.3\pm0.3$ & $[-4,-1]$ & \citet[Tab.\,1]{2003PASP..115..763C} \\
  $\log_{10}\left(N_\mathrm{Ia}\right)$ & Number of SN\,Ia exploding per $\mathrm{M}_\odot$ over 15\,Gyr & $-2.75\pm0.3$ & $[-5,-1]$ & \citet[Tab.1\,]{2012PASA...29..447M}\\
\hline
  \multicolumn{5}{c}{$\vec{\Theta}_i$: \textit{Local ISM parameters}}\\
\hline
  $\log_{10}\left(\mathrm{SFE}\right)$ & Star formation efficiency governing gas infall & $-0.3\pm0.3$ & $[-3,2]$ & \cite{2008AJ....136.2846B}\\
  $\log_{10}\left(\mathrm{SFR}_\mathrm{peak}\right)$ & SFR peak in Gyr (scale of $k=2$ $\Gamma$-distribution)& $0.55\pm0.1$ & $[0.294,1]$ & \citet[fig.\,4b]{2013ApJ...771L..35V}\\
 x$_\mathrm{out}$ & Fraction of stellar feedback outflowing to the gas reservoir & $\phantom{-}0.5\pm0.1$ & $[0,1]$ & \citet[Tab.\,1]{chempy}\\
\hline
 \multicolumn{5}{c}{$T_i$: \textit{Timescale}}\\
\hline
$T_i$ & Time of stellar birth in Gyr & - & [$1$,$13.8$] & Observational Stellar Data
\label{tab:priors}
\end{tabular}
\end{center}
\end{minipage}
\end{table*}
\end{tiny}

To ensure maximal compatibility with TNG, we adopt their nucleosynthetic yield tables in \textit{Chempy}, for enrichment by SN\,Ia, SN\,II and AGB stars. The utilized yields are summarized in Tab.\,\ref{tab:chempy_TNG_yields}, matching \citet[Tab.\,2]{2018MNRAS.473.4077P}, \resub{and we note that the SN\,II yields are renormalized such that the IMF-weighted yield ratios at each metallicity are equal to those from the \citet{2006ApJ...653.1145K} mass range models alone. \textit{Chempy} uses only net yields, such that they provide only newly synthesized material, with the remainder coming from the initial SSP composition.} These tables may not well-represent true stellar chemistry, and the effects of this are examined in Sec.\,\ref{subsec: mocks_wrong_yield} by performing inference using an alternative set of yields. For the analysis of observational data, we would want to use the most up-to-date yields, such as \citet{2016ApJ...825...26K} AGB yields, and carefully chose elements which are known to be well reproduced by our current models (e.g. shown by \citet{2019ApJ...874..102W,2019arXiv190806113G}), though this is not appropriate in our context. To facilitate best comparison with TNG, we further set the maximum SN\,II mass as $100\,\mathrm{M}_\odot$ (matching the IMF upper mass limit), adopt stellar lifetimes from \citet{1998A&A...334..505P} and do not allow for any `hypernovae' (in contrary to \citetalias{2018ApJ...861...40P}).

\begin{table}[]
\caption{Nucleosynthetic yield tables used in this analysis, matching those of the TNG simulation \citep[Tab.\,2]{2018MNRAS.473.4077P}.}
    \centering
    \begin{tabular}{c|c}
      Type & Yield Table \\
       \hline
        SN\,Ia & \citet{1997NuPhA.621..467N}\\
        SN\,II & \citet{2006ApJ...653.1145K,portinari}\\
        AGB & \citet{2010MNRAS.403.1413K,2014MNRAS.437..195D};\\
        & \citet{2014ApJ...797...44F}
    \end{tabular}
\label{tab:chempy_TNG_yields}
\end{table}

TNG only tracks nine elements in their analysis: C, Fe, H, He, Mg, N, Ne, O and Si, reporting the mass-fractions of each \citep{2018MNRAS.473.4077P}. In our analysis we principally compare the logarithmic abundances [X/Fe] and [Fe/H] (defined by 
\begin{equation}
    [\mathrm{X}/\mathrm{Y}] = \log_{10}(N_\mathrm X/N_\mathrm Y)_\mathrm{star} - \log_{10}(N_\mathrm X/N_\mathrm Y)_\odot
\end{equation}
for number fraction $N_\mathrm X$ of element X), where $\odot$ denotes the solar number fractions of \citet{2009ARA&A..47..481A}. This uses H for normalization, thus we are left with $n_\mathrm{el}=8$ independent elements which must be tracked by \textit{Chempy}.\footnote{In observational contexts, it may be more appropriate to compute abundances relative to Mg rather than Fe (as in \citealt{2019ApJ...874..102W}) since Mg is only significantly produced by SN\,II and hence a simpler tracer of chemical enrichment.} 
In this paper, \textit{Chempy} will be used as the principal GCE model, which, with the modifications described above, allows for fast prediction of TNG-like chemical abundances for a given set of galactic parameters. It is important to note that the two GCE models have very different parametrizations of galactic physics, with TNG including vastly more effects, thus it is not certain \textit{a priori} how useful \textit{Chempy} will be in emulating the TNG simulation, although its utility was partially demonstrated in \citetalias{2018ApJ...861...40P}. This is a necessary test to prepare for an inference on real data.

\section{Neural Networks}\label{sec: neural_nets}
Despite the simplifications made by emulating the TNG simulations with the simple GCE model \textit{Chempy}, we will still have difficulties sampling the distribution of the global parameters $\vec\Lambda = \{\alpha_\mathrm{IMF},\log_{10}(N_\mathrm{Ia})\}$ due to the run-time of \textit{Chempy} and the high-dimensionality of the parameter space. To ameliorate this, we utilize \textit{neural networks}; fast non-linear functions containing a large number of trainable parameters.

According to the `Universal Approximation Theorem' \citep{csaji2001approximation}, an arbitrarily complex smooth function can be approximated to any given level of precision by a feed-forward neural network with a finite number of `neurons' ($n_\mathrm{neuron}$) and a single-hidden layer, practically acting as a non-linear interpolator. This implies that, given sufficient training data, a neural network can represent the \textit{Chempy} function arbitrarily well. In essence, instead of computing the full model for each input parameter set, we pass the parameters to the network which predicts the output abundances to high accuracy. This has two benefits;
\begin{enumerate}
    \item \textbf{Speed:} The run-time of the \textit{Chempy} function is $\sim1$\,s per input parameter set, which leads to very slow posterior sampling. With the neural network, this reduces to $\sim5\times10^{-5}$\,s, and is trivially parallelizable, unlike \textit{Chempy}.
    \item \textbf{Differentiablility:} The neural network has a simple closed-form analytic structure (described in appendix \ref{appen: neural_network}), unlike the complex \textit{Chempy} model. This allows it to be differentiated, so we can sample via advanced methods (cf.\,Sec.\,\ref{sec: sampling}).
\end{enumerate}

Despite the additional complexity introduced by using multiple stellar data-points, our network simply needs to predict the birth-time abundances for a single star (with index $i$) given a set of six parameters; $\{\vec\Lambda,\vec\Theta_i,T_i\}$. The same network can be used for all $n_\mathrm{stars}$ stars (and run in parallel), reducing a set of $n_\mathrm{stars}$ runs of \textit{Chempy} to a single matrix computation (with input and output matrices being formed of the stacked parameter and abundance vectors). In this implementation (which differs from that of \citetalias{2018ApJ...861...40P}), we use a sparsely-connected single-layer network with $n_\mathrm{neuron}=40$ neurons for each of $n_\mathrm{el}=8$ abundance outputs. This is trained with a sample of $10^6$ sets of input parameters and output abundances, with hyperparameter optimization and testing performed with an independent sample of consisting of $5\times 10^4$ parameter sets. With the above choices, the network predicts abundances with an average error of $0.005_{-0.004}^{+0.008}$\,dex, which is far below typical observational errors and even smaller away from the extremes of parameter space. Technical details of the network and implementation are discussed in appendix \ref{appen: neural_network}.

\section{The Statistical Model}\label{sec: stat_model}
We here extend the Bayesian model introduced in \citetalias{chempy} to include multiple stellar data-points. Consider a given star with index $i$, born in some region of the ISM. This will carry its own set of parameters $\{\vec\Lambda,\vec\Theta_i,T_i\}$, where $\vec\Lambda$ are taken to be global (hence independent of the stellar label $i$), but the ISM parameters $\vec\Theta_i$ and the birth-time $T_i$ are specific to the star. Using the \textit{Chempy} function (or the trained neural network) we can compute the output $n_\mathrm{el}$ chemical abundances $\{X_i^j\}$ for the $i$-th star as 
\begin{equation}\label{eq: chempy_function}
\{X_i^j\} = \text{Chempy}(\vec\Lambda,\vec\Theta_i,T_i),
\end{equation}
where $j$ indexes the chemical element. These can be compared against observations, with measured abundances $d_i^j$ and corresponding Gaussian errors $\sigma_{i,\mathrm{obs}}^j$, jointly denoted $D_i=\{d_i^j,\sigma_{i,\mathrm{obs}}^j\}$. In addition, we add a star-independent `model error' parameter $\sigma^j_\mathrm{model}$ for each element, which accounts for imperfections in our GCE model (e.g. due to incorrect yields) and is allowed to vary freely.\footnote{This is similar to the model error introduced in \citetalias{2018ApJ...861...40P}, but we now allow it to vary between elements.} This allows the inference to give less weight to elements that are empirically found to fit the data less well. The $i$-th star likelihood is thus simply a product over $n_\mathrm{el}$ Gaussians;
\begin{equation}\label{eq: indiv_likelihood}
    \mathcal{L}_i(D_i|\vec\Lambda,\vec\Theta_i,T_i,\Sigma) = \prod_{j=1}^{n_\mathrm{el}}\frac{1}{\sqrt{2\pi(\sigma^j_{i,\mathrm{tot}})^2}}\exp\left(-\frac{(d_i^j-X_i^j)^2}{2(\sigma_{i,\mathrm{tot}}^j)^2}\right),
\end{equation}
where $\sigma_{i,\mathrm{tot}}^j = \sqrt{(\sigma_{i,\mathrm{obs}}^j)^2+(\sigma_\mathrm{model}^j)^2}$, combining errors in quadrature and denoting the model errors by $\Sigma = \{\sigma_\mathrm{model}^j\}$.

For a collection of $n_\mathrm{stars}$ stellar data-points with the local parameter set $\{\vec\Theta_i\}$ and birth-times $\{T_i\}$, the joint likelihood is simply a product over the individual likelihoods:
\begin{equation}
    \mathcal{L}(\{D_i\}|\vec\Lambda,\{\vec\Theta_i\},\{T_i\},\Sigma) = \prod_{i=1}^{n_\mathrm{stars}}\mathcal{L}_i(D_i|\vec\Lambda,\vec\Theta_i,T_i,\Sigma).
\end{equation}

The full posterior function is derived simply via Bayes rule as
\begin{eqnarray}\label{eq: posteroir}
    \mathbb{P}(\vec\Lambda,\{\vec\Theta_i\},\{T_i\},\Sigma|\{D_i\}) &\propto&  \left[\prod_{i=1}^{n_\mathrm{star}}p_{\vec\Theta}(\vec\Theta_i)p_{T_i}(T_i)\right]\\\nonumber
    &\times& p_{\vec\Lambda}(\vec\Lambda) \times \prod_{j=1}^{n_\mathrm{el}}p_\Sigma(\sigma_\mathrm{model}^j)\\\nonumber
    &\times& \mathcal{L}(\{D_i\}|\vec\Lambda,\{\vec\Theta_i\},\{T_i\},\Sigma)
\end{eqnarray}
where $p_V(V_i)$ is the prior on variable $V_i$ (belonging to the set $V$). The priors are chosen to have the following form:
\begin{itemize}
    \item $\vec\Lambda$: Gaussian priors for $\alpha_\mathrm{IMF}$ and $\log_{10}(N_\mathrm{Ia})$ with parameters defined in Tab.\,\ref{tab:priors}.
    \item $\vec\Theta_i$: Gaussian priors for $\log_{10}(\text{SFE})$ and $\mathrm{x}_\mathrm{out}$ according to Tab.\,\ref{tab:priors} with a truncated Gaussian prior for the peak SFR parameter, restricting to $\log_{10}(\mathrm{SFR}_\mathrm{peak})>0.294$ (cf.\,Sec.\,\ref{subsec: chempy_intro}). Although $\vec\Theta_i$ is different for each star, each vector is taken to be a draw from a star-independent prior.\footnote{A more refined approach would be to assume a full hierarchical structure, where each $\vec\Theta_i$ was a draw from some distribution whose parameters were allowed to vary freely, themselves drawn from a hyperprior, e.g. promoting the mean and variance of $p_\vec{\Theta}$ to be free parameters. This adds additional complexity and is not explored in this paper.} 
    \item $T_i$: Gaussian prior for each star independently. The prior parameters are set from an estimate of the star's birth-time and its variance, representing our best knowledge of this parameter. In experimental contexts, this would be found from age-models (e.g. in the Cannon model \citep{2016ApJ...823..114N} for red giant stars in the APOGEE \citep{2016AN....337..863M} survey).
    \item $\Sigma = \{\sigma_\mathrm{model}^j\}$: Half-Cauchy prior with shape parameter (standard deviation) $\beta_\mathrm{model} = 0.01$. This choice of prior (defined for $\sigma_\mathrm{model}^j\geq0$) allows for arbitrarily small errors, as well as those much greater than the observational errors ($\sim 0.05$\,dex) for poorly reproduced elements.
\end{itemize}

\begin{figure}
\begin{minipage}{\linewidth}
\begin{tikzpicture}

  \node[obs]          (y)   {$d_i^j$}; %
  \factor[above=1 of y] {y-f} {left:$\mathcal{N}$} {} {} ; %
  
  \node[det, above= 0.3of y-f] (X) {$X_i^j$};

  \node[det, above=of X]            (dot) {NN} ; %
  \node[latent, above=0.6 of dot]  (Theta)   {$\mathbf{\Theta}_i$}; %
  \node[latent, right=0.8 of dot] (T)   {$T_i$}; %
  
  \node[const, above=1 of Theta, xshift=-0.5cm] (muTh) {$\mu_\mathbf{\Theta}$} ; %
  \node[const, above=1 of Theta, xshift=0.5cm]  (sigTh) {$\sigma_\mathbf{\Theta}$} ; %
  
  \node[latent, left=0.9 of dot] (Lambda) {$\mathbf{\Lambda}$};
  \node[const, above=1 of Lambda, xshift=-0.5cm] (muL) {$\mu_\mathbf{\Lambda}$};
  \node[const, above=1 of Lambda, xshift=0.5cm] (sigL) {$\sigma_\mathbf{\Lambda}$};

  \node[const, above=1 of T, xshift=-0.5cm] (muT) {$\mu_{T_i}$} ; %
  \node[const, above=1 of T, xshift=0.5cm]  (sigT) {$\sigma_{T_i}$} ; %
  
  \node[obs,right=0.7of y] (ObsErr) {$\sigma^j_{i,\mathrm{obs}}$} ;
  \node[det,right=0.7of y-f] (TotErr) {$\sigma^j_{i,\mathrm{tot}}$};
  
  \node[latent, right=3.cm of y-f] (Err)   {$\sigma_\mathrm{model}$}; %
  \node[const, above = 1.4 of Err](betaErr) {$\beta_\mathrm{model}$};

  \factor[above=of Lambda] {Lambda-f} {left:$\mathcal{N}$}{muL,sigL}{Lambda};
  \factor[above=of T] {T-f} {left:$\mathcal{N}$} {muT,sigT} {T} ; %
  \factor[above=of Theta] {Theta-f} {left:$\mathcal{N}^*$} {muTh,sigTh} {Theta} ; 
  \factor[above=of Err] {Err-f} {left:$\text{H-C}$} {betaErr} {Err} ; %
  \factoredge {X,TotErr} {y-f} {y} ; %

  \edge[->] {T,Theta,Lambda} {dot} ;
  \edge[->] {dot} {X};
  \edge[->] {ObsErr,Err} {TotErr};

  \plate {yx} { %
    (y)(y-f)
    (muT)
    (sigT)
    (Theta)(Theta-f)
    (T)(T-f)
    (X) %
    (ObsErr)
    (TotErr)
    (dot) %
  } {$n_\mathrm{stars}$} ;
  \plate {} {%
    (y)(y-f)
    (X) %
    (Err)(Err-f)
    (yx.south west)
  } {$n_\mathrm{el}$} ;

\end{tikzpicture}
\end{minipage}
\caption{Probabilistic Graphical Model for the statistical inference used in this paper. Unfilled circular nodes, filled circular nodes and diamond nodes represent random variables, observed data and deterministic calculations respectively. Prior parameters (such as $\mu_\vec{\Lambda}$) are shown without nodes and the boxes indicate how many of each feature are present (e.g. there are $n_\mathrm{stars}$ $\vec\Theta_i$ realizations). Parameters outside the boxes have only a single value (independent of the element and star analyzed). NN represents the neural network (or the \textit{Chempy} function), which produces output abundances $X_i^j$ for element $j$ in star $i$. Here, $\mathcal{N}^*$ and H-C represent (possibly truncated) Normal and Half-Cauchy prior distributions. Figure created using TikZ Bayesnet.}\label{fig: overall_pgm}
\end{figure}
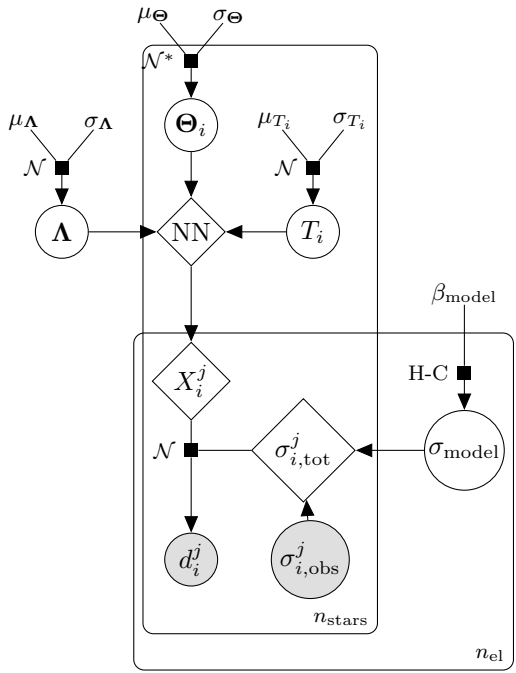

In statistical language, the model can be expressed as 
\begin{eqnarray}\label{eq: statistical_model}
    \vec{\Lambda}&\sim&p_{\vec\Lambda}=\mathcal{N}(\vec\mu_\Lambda,\vec\sigma_\Lambda)\\\nonumber
    \vec{\Theta}_i&\sim&p_{\vec\Theta}=\mathcal{N}^{*}(\vec\mu_\Theta,\vec\sigma_\Theta)\\\nonumber
    T_i&\sim&p_{T_i}=\mathcal{N}(\mu_{T_i},\sigma_{T_i})\\\nonumber
    \sigma_\mathrm{model}^j&\sim&p_\Sigma = \text{Half-Cauchy}(\beta_\mathrm{model})\\\nonumber
    \{X_i^j\} &=& \text{Chempy}(\vec\Lambda,\vec\Theta_i,T_i)\\\nonumber
    \sigma_{i,\mathrm{tot}}^j&=&\sqrt{(\sigma_{i,\mathrm{obs}}^j)^2+(\sigma_\mathrm{model}^j)^2}\\\nonumber
    X_i^j &\sim& \mathcal{N}(d_i^j,\sigma_{i,\mathrm{tot}}^j)
\end{eqnarray}
where $\mathcal{N}^{*}$ indicates a possibly truncated Gaussian (for the SFR parameter). In total, we have $2+4n_\mathrm{stars}+n_\mathrm{el}$ free parameters to be inferred from $n_\mathrm{el}n_\mathrm{stars}$ data-points, given $6+n_\mathrm{stars}$ individual prior distributions. This is summarized in Fig.\,\ref{fig: overall_pgm}, in the form of a Probabilistic Graphical Model (PGM), which shows the relationship between all variables and hyperparameters.

\section{Sampling Techniques}\label{sec: sampling}
To determine the optimal values of the global galactic parameters ($\vec\Lambda$) we must sample the posterior of Eq.\,\ref{eq: posteroir}. In previous work \citepalias{chempy,2018ApJ...861...40P}, this was acheived using Ensemble Sampling Markov Chain Monte Carlo (MCMC) using the \texttt{emcee} package \citep{2013PASP..125..306F}. The authors of \texttt{emcee} note that this is not appropriate for sampling high-dimensional parameter spaces, thus here, where the dimensionality scales with $n_\mathrm{stars}$, we must find an alternative sampler. Gibbs sampling \citep{Gibbs} is one option, where marginal posterior functions are used to iteratively first update the global $\vec\Lambda$ and $\Sigma$ parameters and then the local $\{\vec\Theta_i,T_i\}$ parameters, based on a Metropolis-Hastings sampling approach \citep{MetropolisHastings}. However, this is difficult to use in practice, due to (a) the requirement of knowing the marginal posterior functions (e.g. $\mathbb{P}(\vec\Lambda|\{\vec\Theta_i,T_i\},\Sigma, \{D_i\})$), (b) the large number of tunable parameters, and (c) slow convergence. 

Here we principally consider the modern sampling technique `Hamiltonian Monte Carlo' \citep[HMC;][]{2012arXiv1206.1901N}, which uses posterior function gradients to sample much more efficiently than canonical MCMC methods. This can also sample much higher-dimensional posteriors than Ensemble Sampling. The basic premise (explained in more detail in appendix \ref{appen: HMC_details}) is as follows. In standard MCMC approaches, given a current position in the MCMC chain, the next position is chosen via a random jump such that the chain traverses a random walk in parameter space. By introducing additional `momentum' parameters, we can choose samples in a more efficient manner akin to a rocket exploring the space around a planet by traversing orbits of constant energy then making random jumps in energy rather than just jumping between positions at random. This however requires the posterior function to be differentiable, which is seldom possible for complex astronomical models. In this context, the replacement of \textit{Chempy} by a trained neural network gives a trivially differentiable model, since the network is a simple function of matrices and $\tanh$ functions, thus HMC can be used in our context.

In practice, this is implemented using the Python \texttt{PyMC3} package \citep{pymc3},\footnote{\href{https://docs.pymc.io/}{docs.pymc.io/}} utilizing the `automatic differentiation' routines from \texttt{theano} \citep{theano} to compute the posterior gradients. HMC sampling is performed via the `No U-Turn Sampler' \citep[NUTS; ][]{2011arXiv1111.4246H} using $16,000$ chain samples with a desired sample-acceptance rate of $0.9$. \resub{The sampler uses $2\times 10^4$ initialization steps (which set the start point of the Markov chain) and 2000 `tuning' steps (to adjust internal parameters and stabilize the Markov chain), with sampling expedited by running multiple smaller chains in parallel on different CPUs, which can then be combined.} 


In the case of very large $n_\mathrm{stars}$, the dimensionality of our problem becomes large, and we find that even HMC requires an unwieldy sampling time. For this reason, we restrict to $n_\mathrm{stars}\lesssim200$ for the HMC analysis to ensure sampling can be done in a few tens of CPU-hours. For a larger sample of stars, we may look to approximate sampling methods, such as `Automatic Differentiation Variational Inference' \citep[ADVI; ][]{2013arXiv1312.6114K,2016arXiv160300788K,2017arXiv170309194R}. \resub{This is briefly discussed in appendix \ref{appen: HMC_details}, and a simple form (Mean Field ADVI) is used for the NUTS initialization steps.} Using $n_\mathrm{stars}=200$ is found to give well-constrained posterior parameter estimates in this paper, thus we do not implement ADVI here.


\section{Results}\label{sec: results}
In this section we apply the statistical techniques described in Sec.\,\ref{sec: sampling} to the posteriors of Sec.\,\ref{sec: stat_model} to infer the global galactic parameters $\alpha_\mathrm{IMF}$ and $\log_{10}(N_\mathrm{Ia})$. To demonstrate the utility of our method, we compare the derived global parameters with the true values, using three mock data-sets:
\begin{enumerate}
    \item A data-set created by \textit{Chempy} with the same nucleosynthetic yield tables as for the neural network training. This is used to test the sampling techniques and neural networks;
    \item A data-set created by \textit{Chempy} with different yield tables to that of the neural network. This is used to test the dependence of our inference on the yield tables;
    \item A data-set derived from stellar particles taken from a galaxy in the TNG simulation (yields are the same as in case 1). This is used to test the dependence of our inference on the galactic physics parametrization.
\end{enumerate}

In each case we obtain a set of stellar birth-times and chemical abundances, that, to fully represent observational data, must be augmented with errors. In line with typical APOGEE \citep{2016AN....337..863M} abundance data, we conservatively assume \resub{a uniform Gaussian error of 0.05\,dex in the [Fe/H] and [X/Fe] values}. In addition, we assign a 20\% fractional error to each birth-time measurement $\{T_i\}$, roughly matching that obtained in current analyses using APOGEE data \citep{2016ApJ...823..114N}. Mock `observed' abundances and birth-times are drawn from Gaussian distributions about their true values with the above errors and we disregard any stars with `observed' birth-times (i.e. the prior means) $\mu_{T_i}\notin [1,13.8]\,$Gyr. The outcome of this mock data creation is a set of 200 mock stars, all with relevant observational abundances and birth-times, emulating a real data-set. These data-sets have been made freely available online alongside a tutorial showing their format and usage.\footnote{\href{https://github.com/oliverphilcox/ChempyMulti}{github.com/oliverphilcox/ChempyMulti}}

\subsection{Mock Data from \textit{Chempy}}\label{subsec: mocks_Chempy}

To create the \textit{Chempy} mock data, we first set the values of the global galactic parameters as $\alpha_\mathrm{IMF} = -2.3$ and $\log_{10}(N_\mathrm{Ia}) = -2.89$, matching those used by TNG \citep{2018MNRAS.473.4077P}. Using the priors in Tab.\,\ref{tab:priors}, we then create a set of 200 random draws of the local parameters $\vec\Theta_i = \{\log_{10}(\text{SFE}),\log_{10}(\text{SFR}_\mathrm{peak}),\text{x}_\mathrm{out}\}$, additionally drawing $T_i$ uniformly from the range $[2,12.8]$\,Gyr, to minimize overlap with the neural network training birth-time limits when observational uncertainties are included.\footnote{We note that the choice of stellar age distribution is unimportant here, as long as all birth-times are inside the neural network training limits.} Each set of parameters is passed to the \textit{Chempy} function, producing eight output \textit{true} chemical element abundances that are then augmented with errors, as above.

Following this, the methods of Sec.\,\ref{sec: sampling} are used to infer the posterior distribution of $\vec\Lambda$ by sampling the full high-dimensional parameter space via the HMC algorithm. Here, \textit{Chempy} is being used both to create and fit the data, thus there is no mismatch between observations and sampler in terms of physics parametrization or yield tables. This should imply small model errors (i.e. $\sigma_\mathrm{model}^j\rightarrow0$), though the model errors are retained in the inference as a useful test. Analysis is performed for a selection of $n_\mathrm{stars}\in[1,200]$. To illustrate the bias created by using only a small selection of stars, we split a sample of 200 stars into non-intersecting sub-samples of size $n_\mathrm{stars}$ and perform the inference separately on each (i.e. we perform 100 1-star analyses, 50 2-star analyses etc.). In our implementation (utilizing parallel sampling across 16 cores), the analysis of each sub-sample has a run-time ranging from $\sim  1$\,CPU-minute (for $n_\mathrm{stars}=1$) to $\sim 40$\,CPU-hours (for $n_\mathrm{stars}=200$) on a modern machine.

\begin{table}
\caption{Constraints on the global galactic parameters from Hamiltonian Monte Carlo (HMC) sampling using the three mock data-sets described in Sec.\,\ref{sec: results}. These are also displayed graphically in Fig.\,\ref{fig: mock_comparison}. We state the median posterior estimates for a variety of $n_\mathrm{stars}$ values, taking the median over all independent sub-samples of this size. `Stat.' refers to the median $1\sigma$ posterior distribution width for a single realization (showing the precision possible in a typical measurement) and `Sample' gives the $1\sigma$ variation between sub-samples (illustrating the bias caused by the specific choice of stars in the sub-sample). \resub{The true parameter values are $\alpha_\mathrm{IMF}=-2.3$ and $\log_{10}(N_\mathrm{Ia}) = -2.89$.}}
\begin{tabular}{ c|ccc|ccc }
$n_\mathrm{stars}$ & $\alpha_{\mathrm{IMF}}$ & Stat. & Sample & $\log_{10}(N_\mathrm{Ia})$ & Stat. & Sample\\
 \hline
  \multicolumn{7}{c}{(a) \textit{Chempy} mock data with correct yield set}\\
\hline
$1$ &
 $-2.29$ & $^{+0.08}_{-0.08}$ & $^{+0.07}_{-0.06}$
&
 $-2.87$ & $^{+0.11}_{-0.11}$ & $^{+0.08}_{-0.08}$
 \\
$10$ &
 $-2.31$ & $^{+0.02}_{-0.02}$ & $^{+0.03}_{-0.02}$
&
 $-2.90$ & $^{+0.03}_{-0.03}$ & $^{+0.04}_{-0.02}$
 \\
$100$ &
 $-2.31$ & $^{+0.01}_{-0.01}$ & $^{+0.00}_{-0.00}$
&
 $-2.90$ & $^{+0.01}_{-0.01}$ & $^{+0.00}_{-0.00}$ \\
 \hline
  \multicolumn{7}{c}{(b) \textit{Chempy} mock data with incorrect yield set}\\
\hline
$1$ &
 $-2.25$ & $^{+0.11}_{-0.09}$ & $^{+0.09}_{-0.07}$
&
 $-3.01$ & $^{+0.15}_{-0.15}$ & $^{+0.13}_{-0.11}$
 \\
$10$ &
 $-2.21$ & $^{+0.04}_{-0.04}$ & $^{+0.04}_{-0.05}$
&
 $-2.96$ & $^{+0.08}_{-0.08}$ & $^{+0.05}_{-0.08}$
 \\
$100$ &
 $-2.22$ & $^{+0.02}_{-0.02}$ & $^{+0.01}_{-0.01}$
&
 $-2.96$ & $^{+0.03}_{-0.02}$ & $^{+0.00}_{-0.00}$
 \\
 \hline
  \multicolumn{7}{c}{(c) IllustrisTNG mock data}\\
\hline
$1$ &
 $-2.27$ & $^{+0.08}_{-0.08}$ & $^{+0.15}_{-0.12}$
&
 $-2.86$ & $^{+0.11}_{-0.11}$ & $^{+0.11}_{-0.11}$
 \\
$10$ &
 $-2.27$ & $^{+0.03}_{-0.03}$ & $^{+0.03}_{-0.03}$
&
 $-2.87$ & $^{+0.03}_{-0.04}$ & $^{+0.02}_{-0.02}$
 \\
$100$ &
 $-2.28$ & $^{+0.01}_{-0.01}$ & $^{+0.01}_{-0.01}$
&
 $-2.89$ & $^{+0.01}_{-0.01}$ & $^{+0.00}_{-0.00}$
\label{tab:results}
\end{tabular}
\end{table}

\begin{figure*}[!th]
\centering
    \subfloat[][Inference with \textit{Chempy} mock data with the correct yield set (Sec.\,\ref{subsec: mocks_Chempy})]{\includegraphics[height=0.22\textheight,width=\textwidth]{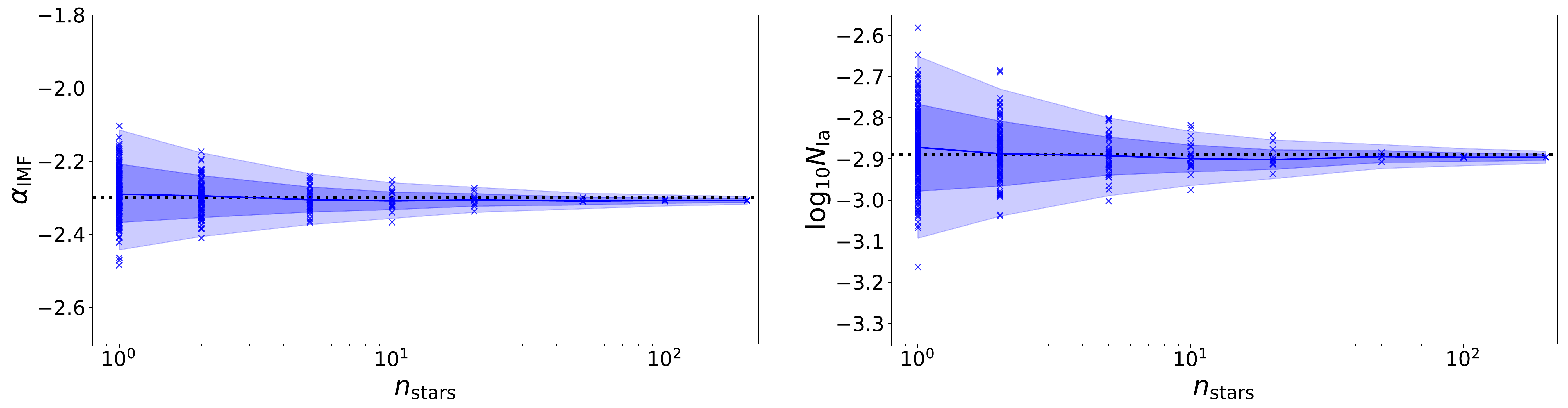} }\\
    \subfloat[][Inference with \textit{Chempy} mock data with an incorrect yield set (Sec.\,\ref{subsec: mocks_wrong_yield})]{\includegraphics[height=0.22\textheight,width=\textwidth]{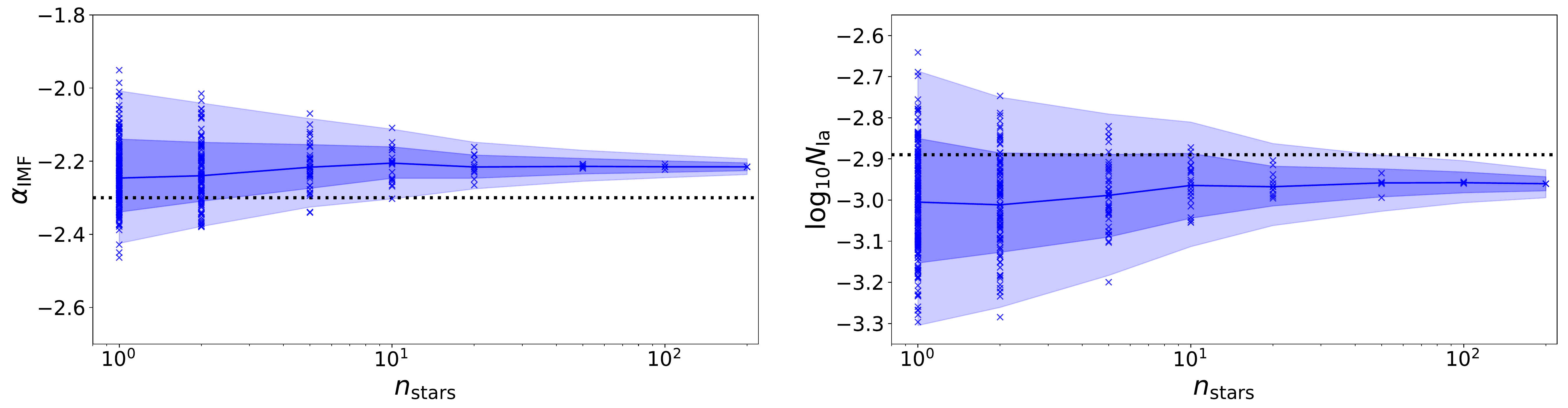} }\\
    \subfloat[][Inference with mock data drawn from an IllustrisTNG Milky Way-like galaxy (Sec.\,\ref{subsec: mocks_TNG})]{\includegraphics[height=0.22\textheight,width=\textwidth]{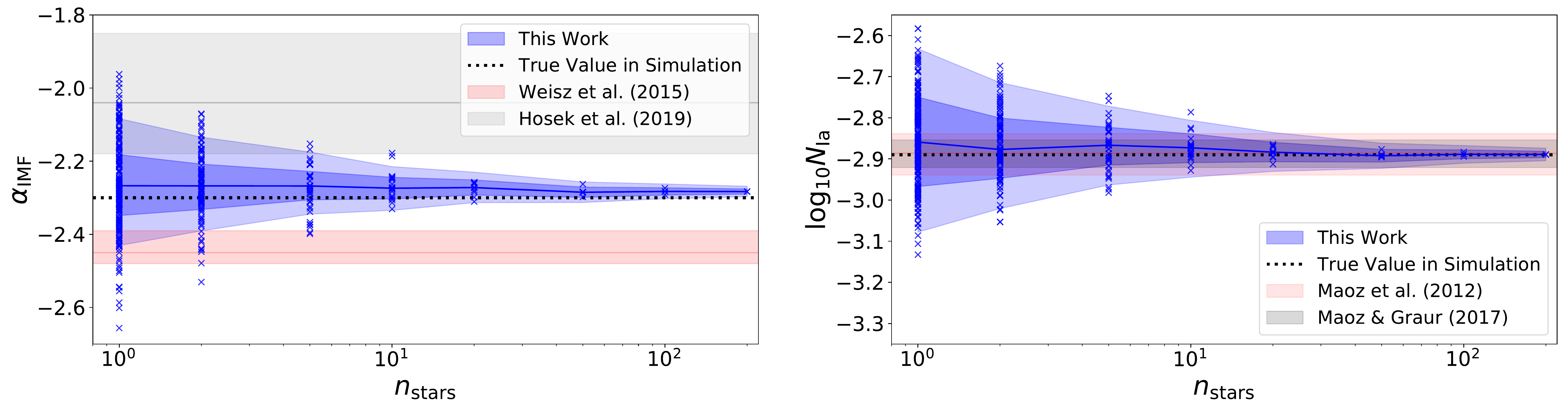} }%
\caption{Posterior bounds on the global parameters $\alpha_\mathrm{IMF}$ (left) and $\log_{10}(N_\mathrm{Ia})$ (right) for three mock data-sets as a function of the number of stars in the sample, $n_\mathrm{stars}$. Blue data-points represent the median parameter estimate \resub{for each disjoint subset of the full sample at fixed $n_\mathrm{stars}$}, with a solid line giving the median value across all sub-samples. Dark (light) filled blue regions indicate the $1\sigma$ ($2\sigma$) \textit{statistical} uncertainty obtained from a single sub-sample of $n_\mathrm{stars}$, taking the median across all realizations. There is an additional \textit{sample} variance caused by only using a small number of stars in the analysis, shown by the variation of parameter medians across sub-samples at fixed $n_\mathrm{stars}$. A dotted line indicates the true global parameter values and all inference is performed via Hamiltonian Monte Carlo (HMC) sampling. For context, in (c) we additionally show $\alpha_\mathrm{IMF}$ bounds from star counts in M31 \citep{2015ApJ...806..198W} and the Milky Way \citep{2019ApJ...870...44H}, as well as $\log_{10}(N_\mathrm{Ia})$ constraints from \citet{2012MNRAS.426.3282M} and \citet{2017ApJ...848...25M}. \resub{Since} the results in this paper are with reference to simulated data only we do not expect agreement in the inferred parameter medians. \resub{For (a) and (c), the parameters appear to converge} to the true values as $n_\mathrm{stars}$ becomes large, with some bias seen in \resub{(b)}.}
    \label{fig: mock_comparison}%
\end{figure*}

\begin{table}
\caption{Inferred model error parameters, $\sigma^j_\mathrm{model}$, from HMC sampling using the three mock data-sets of Sec.\,\ref{sec: results} and three values of $n_\mathrm{stars}$. These show the how well each element is reproduced by the \textit{Chempy} model (with lower errors implying a smaller model discrepancies), and are added to observational errors in quadrature. For each, we show the median and $1\sigma$ parameter constraints for three representative elements (averaged over all sub-samples at fixed $n_\mathrm{stars}$), with the full distributions for $n_\mathrm{stars}=200$ being shown in Fig.\,\ref{fig: model_errors}. \resub{The prior is given by $\sigma_\mathrm{model}^j =  0.010_{-0.007}^{+0.030}$. Corresponding} posterior constraints on the global parameters are shown in Tab.\,\ref{tab:results}.}
\begin{tabular}{ c|cc|cc|cc }
$n_\mathrm{stars}$ & \multicolumn{2}{c}{[Fe/H]} & \multicolumn{2}{c}{[C/Fe]} & \multicolumn{2}{c}{[N/Fe]} \\
 \hline
  \multicolumn{7}{c}{(a) \textit{Chempy} mock data with correct yield set}\\
\hline
$1$ & $0.009$ & $^{+0.021}_{-0.007}$
& $0.009$ & $^{+0.020}_{-0.007}$
& $0.009$ & $^{+0.021}_{-0.007}$
 \\
$10$ &
 $0.008$ & $^{+0.014}_{-0.006}$
& $0.008$ & $^{+0.014}_{-0.006}$
& $0.007$ & $^{+0.011}_{-0.005}$
 \\
$100$ & $0.006$ & $^{+0.009}_{-0.005}$
& $0.007$ & $^{+0.009}_{-0.005}$
& $0.005$ & $^{+0.007}_{-0.004}$
\\
 \hline
  \multicolumn{7}{c}{(b) \textit{Chempy} mock data with incorrect yield set}\\
\hline
$1$ & $0.009$ & $^{+0.022}_{-0.007}$
& $0.170$ & $^{+0.193}_{-0.089}$
& $0.014$ & $^{+0.060}_{-0.010}$
 \\
$10$ & $0.008$ & $^{+0.014}_{-0.006}$
& $0.268$ & $^{+0.074}_{-0.053}$
& $0.141$ & $^{+0.049}_{-0.036}$
 \\
$100$ & $0.006$ & $^{+0.009}_{-0.004}$
& $0.265$ & $^{+0.022}_{-0.020}$
& $0.159$ & $^{+0.015}_{-0.014}$
 \\
 \hline
  \multicolumn{7}{c}{(c) IllustrisTNG mock data}\\
\hline
$1$ & $0.009$ & $^{+0.022}_{-0.007}$
& $0.009$ & $^{+0.021}_{-0.007}$
& $0.009$ & $^{+0.024}_{-0.007}$
 \\
$10$ & $0.020$ & $^{+0.072}_{-0.016}$
& $0.009$ & $^{+0.017}_{-0.007}$
& $0.008$ & $^{+0.014}_{-0.006}$
 \\
$100$ & $0.217$ & $^{+0.022}_{-0.021}$
& $0.017$ & $^{+0.010}_{-0.010}$
& $0.005$ & $^{+0.007}_{-0.004}$
\label{tab:model_errors}
\end{tabular}
\end{table}

The resulting posterior distribution parameters of $\vec\Lambda$ are summarized in Fig.\,\ref{fig: mock_comparison}a and Tab.\,\ref{tab:results}a. For the measurement of global parameters in a sub-sample of stars we note two contributions to the variance; (a) the intrinsic \textit{statistical} variance from the width of the posterior distribution for $\vec\Lambda$ (shown by the shaded regions in the plot), and (b) the \textit{sample} variance arising from the bias caused by analyzing only a small set of stars (shown by the spread of individual posterior medians in the plot). For small $n_\mathrm{stars}$, the effects have similar magnitude, with sample variance contributing $\sim 4\%$ to the total uncertainty of each realization for $n_\mathrm{stars}=1$ (quantified by the standard deviation of the median posterior parameter estimates between sub-samples). For large sub-samples, where we include stars from a large variety of ISM environments, the effect is however subdominant. This implies that measuring galactic parameters from a single star can give significantly biased results, which is important to take into account when considering single-star analyses such as \citetalias{chempy}.

Considering the average over all sub-samples at fixed $n_\mathrm{stars}$ (as in Tab.\,\ref{tab:results}a), the median of the posterior inferences are seen to be in full agreement with the true values in all cases, given the statistical errors. For $n_\mathrm{stars}\gtrsim5$ this is additionally true for the estimates from individual sub-samples, confirming that the sample variance effect is of only minor importance at large $n_\mathrm{stars}$. As expected, the statistical widths of the posterior distributions shrink as $n_\mathrm{stars}$ increases, since the number of individual data-points (here $n_\mathrm{el}n_\mathrm{stars}=8n_\mathrm{stars}$) becomes large compared to the number of free parameters $(2+n_\mathrm{el}+4n_\mathrm{stars}=10+4n_\mathrm{stars}$). For $n_\mathrm{stars}=200$ we obtain bounds of $\alpha_\mathrm{IMF}=-2.31\pm 0.01$, $\log_{10}(N_\mathrm{Ia}) = -2.90\pm 0.01$, which is fully consistent, as before.\footnote{Since we only use a single sub-sample for the $n_\mathrm{stars}=200$ analysis, the sample variance cannot be determined. Given the general trend with $n_\mathrm{stars}$ however, we expect it to be small.}

\begin{figure*}[htbp]
\centering
    \subfloat[][\textit{Chempy} mock data with the correct yield set (Sec.\,\ref{subsec: mocks_Chempy})]{\includegraphics[width=0.5\linewidth]{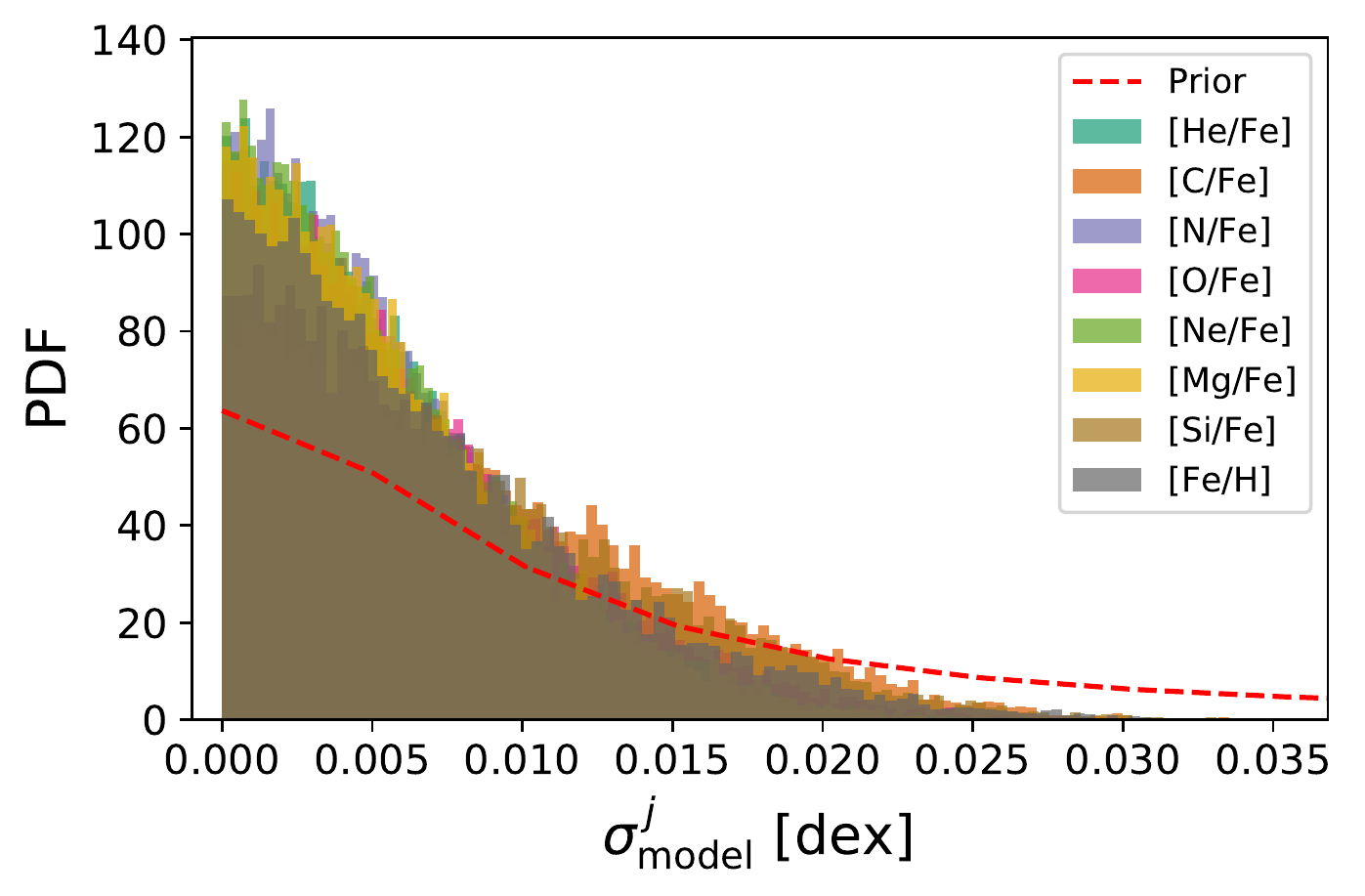}}
    \subfloat[][\textit{Chempy} mock data with an incorrect yield set (Sec.\,\ref{subsec: mocks_wrong_yield})]{\includegraphics[width=0.5\linewidth]{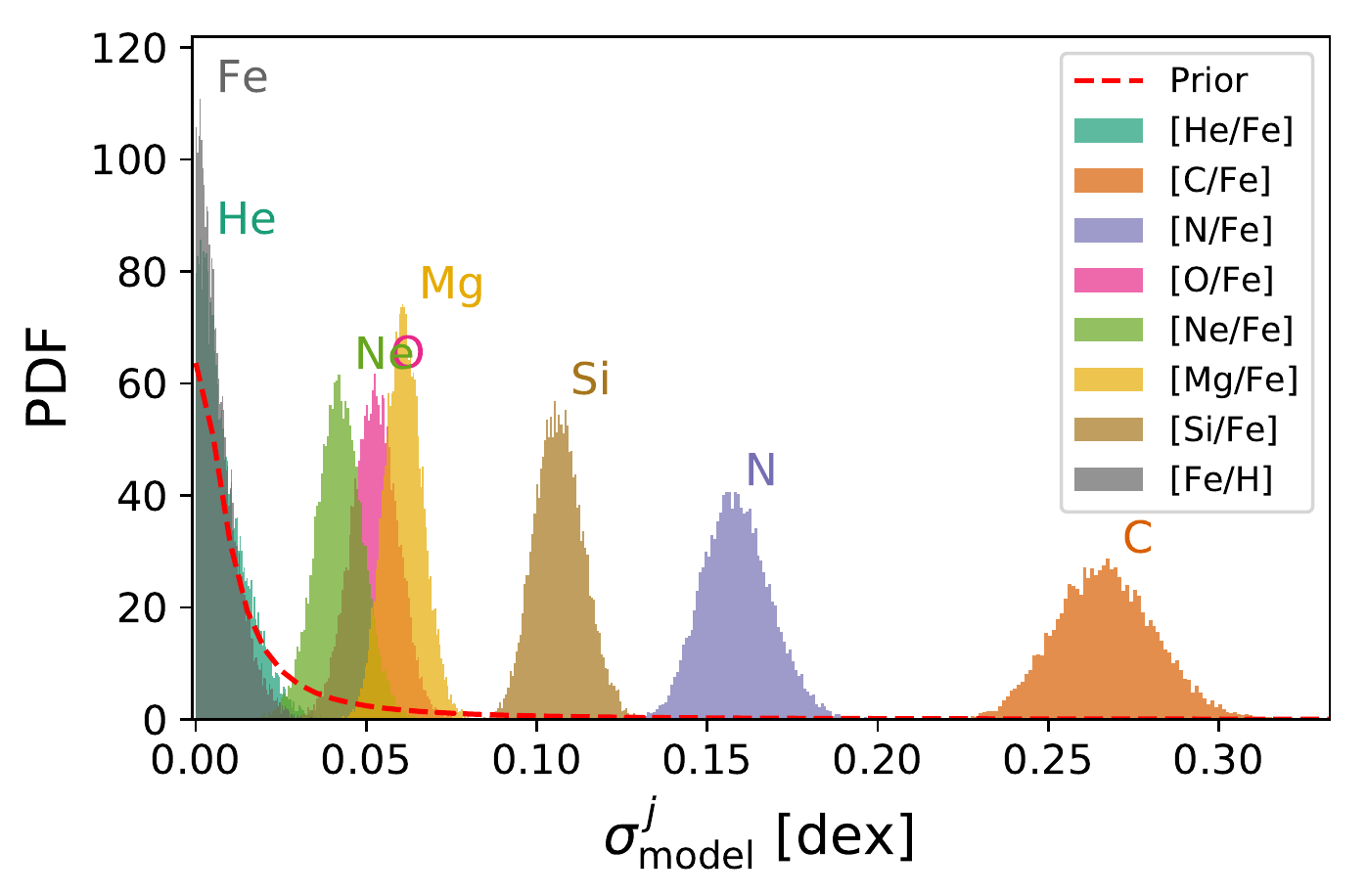} }\\
    \subfloat[][IllustrisTNG mock data (Sec.\,\ref{subsec: mocks_TNG})]{\includegraphics[width=0.5\linewidth]{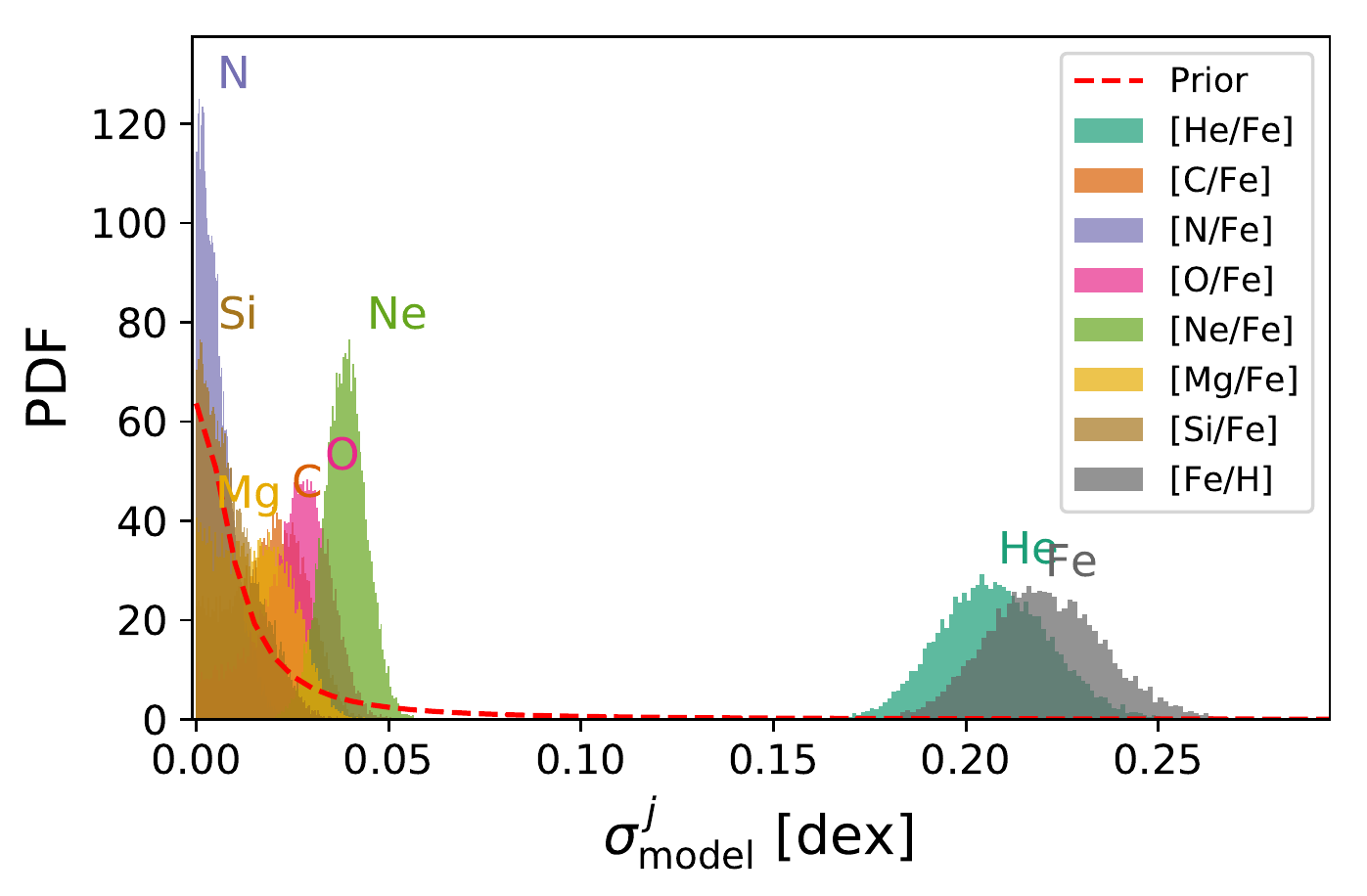} }
    \caption{Posterior distributions of the model error parameters $\Sigma \equiv \{\sigma_\mathrm{model}^j\}$ \resub{obtained from HMC inference using} $n_\mathrm{stars}=200$ and the three data-sets described in Sec.\,\ref{sec: results}. Individual histograms show the results for single elements, with a red dotted line indicating the Half-Cauchy prior assumed. Posterior predictions for the model errors for smaller $n_\mathrm{stars}$ are given in Tab.\,\ref{tab:model_errors}. Note the significantly different $x$-axis ranges between the three plots.}
    \label{fig: model_errors}%
\end{figure*}

Analysis of the posterior model errors, $\Sigma \equiv \{\sigma_\mathrm{model}^j\}$, is performed in Fig.\,\ref{fig: model_errors}a, showing the full posterior distributions for $n_\mathrm{stars}=200$, and Tab.\,\ref{tab:model_errors}a, summarizing the inferred parameters for a range of data-set sizes. We firstly note the model errors to be approximately independent of the element label $j$, as predicted. (We expect all elements to be equally reliable as there is no mismatch between data and sampling model). The distributions are clearly centered on zero, and are similar in form to the priors (Half-Cauchy distributions with standard deviation $\beta = 0.01$\,dex) although they become sharper as $n_\mathrm{stars}$ increases. Taking the median across all elements and sub-samples at fixed $n_\mathrm{stars}$, the average standard deviation of $\sigma_\mathrm{model}^j$ falls from $\approx 0.05$\,dex at $n_\mathrm{stars}=1$ to $\approx 0.005$\,dex at $n_\mathrm{stars}=200$, significantly below the prior value. As $n_\mathrm{stars}$ increases, so does the number of independent data-points, leading to smaller statistical error and hence a reduced standard deviation (given that the prior is peaked at zero). This behavior is fully consistent with the $\sigma_\mathrm{model}^j\rightarrow0$ limit, with no preference shown for non-zero model errors.

We may also consider the constraints that may be placed on the stellar birth-times from this analysis. The posterior estimates of $T_i$ are highly consistent with the true values, with a fractional deviation of $-0.02^{+0.16}_{-0.15}$ ($0.00\pm0.15$) for $n_\mathrm{stars}=1$ ($n_\mathrm{stars}=200$), averaging across all 200 stars. In addition, the posterior distributions are somewhat narrower than the priors, with fractional widths of $0.16_{-0.02}^{+0.01}$ ($0.14\pm0.02$) for $n_\mathrm{stars}=1$ ($n_\mathrm{stars}=200$), compared to the prior width of $20\%$. These constraints are far weaker than those of the global parameters, showing little variation with the sub-sample size. This is because the birth-times belong to the set of local variables (along with the three ISM parameters), which must be constrained by only $n_\mathrm{el}=8$ data-points, unlike the global parameters, which are constrained by all $n_\mathrm{stars}n_\mathrm{el}$ abundances. For larger $n_\mathrm{stars}$, each individual data-point has less effect on $\vec\Lambda$, thus the constraining power of the data on the local parameters increases slightly, though we are still limited by $n_\mathrm{el}$. To obtain sharper constraints, we need only increase the number of elements analyzed. \resub{In applications of this method to observational data, our age analysis would be aided by models of surface chemical abundance change \citep[e.g.][]{2016MNRAS.456.3655M}, as well as implementation of more nucleosynthetic processes, in order to provide age-sensitive elements \citep{2016A&A...593A..65N, 2018MNRAS.474.2580S, 2019A&A...622A..59T}, though in the context of GCE models this usually depends on the galactic component under investigation \citep[e.g.][]{2011A&A...530A..15N,2011ApJ...729...16K}.}

From the above, it is clear that the latter part of our analysis works as expected, with the sampler able to correctly (and precisely) infer global parameters from data which uses the same physical model and yield tables, despite only placing weak constraints on the local parameters. By increasing the number of stars (or the number of chemical elements), we can obtain tighter bounds on global parameters and reduce bias caused by the choice of sub-sample. At this stage however, it is not clear whether this will extend to samples drawn from simulations (or universes) that do not obey the same evolutionary model as \textit{Chempy}.

\subsection{Mock Data with an Incorrect Yield Set}\label{subsec: mocks_wrong_yield}
In the real universe, the chemical yields from stellar nucleosynthetic processes will not exactly match those tabulated in our yield tables (Tab.\,\ref{tab:chempy_TNG_yields}). To investigate the effect of this we consider an analysis using mock data created again by \textit{Chempy}, but with a different set of nucleosynthetic yields. 

\begin{table}[]
\caption{Alternative nucleosynthetic yield tables used in the analysis of Sec.\,\ref{subsec: mocks_wrong_yield} to investigate the effects of incomplete knowledge of the true yield tables on the inferred galactic parameters. These exhibit moderate differences from the yields of Tab.\,\ref{tab:chempy_TNG_yields} as shown graphically in Fig.\,\ref{fig: wrong_yield_comparison}.}
    \centering
    \begin{tabular}{c|c}
      Type & Yield Table \\
      \hline
        SN\,Ia & \citet{2003NuPhA.718..139T}\\
        SN\,II & \citet{nomoto2013}\\
        AGB & \citet{2016ApJ...825...26K}
        
    \end{tabular}
    \label{tab:other_chempy_yields}
\end{table}

\begin{figure*}[htb]
    \centering
    \includegraphics[width=\linewidth]{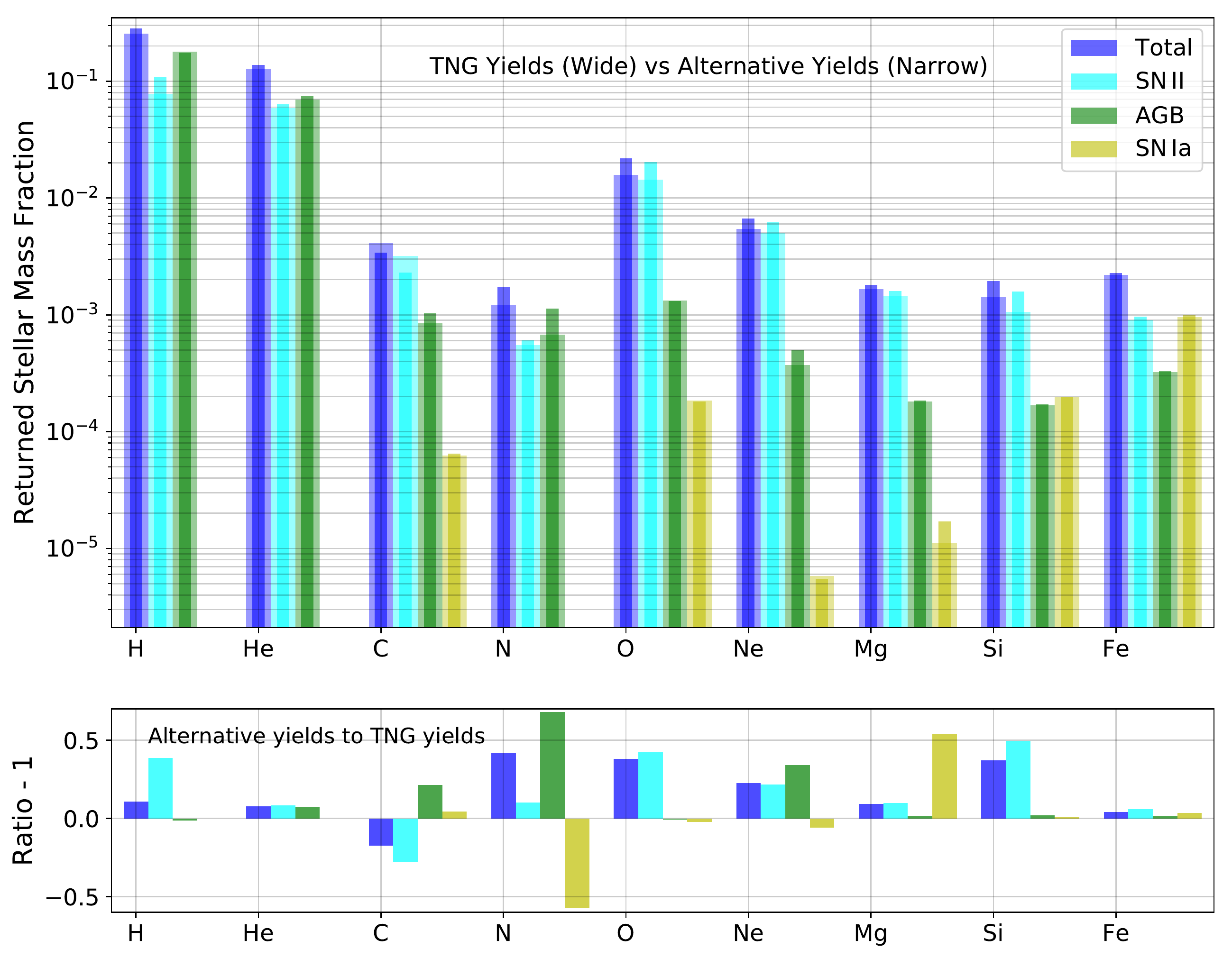}
    \caption{\resub{Mass fraction} returned to the ISM over 13.8\,Gyr for a simple stellar population (SSP) formed at solar metallicity for the eight elements tracked by TNG as well as H (used for abundance normalization). Wide (narrow) bars show the results for TNG (alternative) yield tables described in Tab.\,\ref{tab:chempy_TNG_yields} (Tab.\,\ref{tab:other_chempy_yields}). Both sets of yields are converted from `net' to `gross' form by adding unprocessed mass feedback with element fractions taken from the initial SSP composition (here chosen as solar). The mass return is separated for each tracked nucleosynthetic process and the lower plot shows the fractional difference between the two yield tables (with a linear scale). This figure is analogous to \citet[Fig.\,1]{2018MNRAS.473.4077P}, and we use the same SSP model and yields as TNG.}\label{fig: wrong_yield_comparison}
\end{figure*}

The utilized yield tables are listed in Tab.\,\ref{tab:other_chempy_yields} and have been chosen to \resub{ensure that contributions to all processes differ at $\mathcal{O}(10\%)$}.\footnote{When performing inference with observational data, one would restrict to elements which are known to be well reproduced by current models, avoiding large mismatches between predicted and true yields. For this reason we do not simply use the most up-to-date yield tables here, since, for some elements, they differ from the (older) TNG yields by several orders of magnitude giving a large bias, exceeding that which would be expected in a typical analysis.} In Fig.\,\ref{fig: wrong_yield_comparison}, we visualize both yield sets, plotting the \resub{fractional} mass returned to the ISM by each nucleosynthetic process over $13.8$\,Gyr for \resub{an SSP} formed at solar metallicity. \resub{The mean deviation between the yield sets is $\sim 20\%$, both for the total mass return and for that from the individual nucleosynthetic processes.} The greatest differences are for N, \resub{with a $\sim60\%$ shift in the dominant (AGB) nucleosynthetic channel}, although we also note large changes to the total yield for O and Si \resub{(around 40\%)}. There is additionally a slight increase in the Fe yield for the new yields relative to TNG, which will affect all [X/Fe] abundances via the normalization.\footnote{In principle, this could be ameliorated by performing inference using the metal mass fractions themselves rather than the abundances. The advantage of our approach is that most abundances are insensitive to the metallicity of the star (except for [Fe/H] and [He/Fe]) since they depend only on metal mass ratios.}

Using these yields, mock data were constructed using \textit{Chempy} as in Sec.\,\ref{subsec: mocks_Chempy} and HMC inference performed with the same neural network as before (which was trained with the original TNG yields). Data is thus created with the alternative yield set, but analyzed assuming TNG yields, allowing us to explore the impact of incorrectly assumed yield tables on the output parameters distributions. 

The inference results are summarized in Fig.\,\ref{fig: mock_comparison}b and Tab.\,\ref{tab:results}b, in the same manner as above. Like before, the sample and statistical variances are seen to decrease as a function of $n_\mathrm{stars}$, though we note larger variances in all cases, since the data are less constraining (due to mismatches between observations and model that increase the model error and thus decrease the constraining power). Notably, for $n_\mathrm{stars}\gtrsim 50$, the posterior parameter distributions become \textit{inconsistent} with the true values, with 68\% confidence intervals of $\widehat{\alpha}_\mathrm{IMF}=-2.22\pm0.01$ and $\widehat{\log_{10}(N_\mathrm{Ia})}=-2.96\pm0.02$ obtained for $n_\mathrm{stars}=200$ (ignoring greatly subdominant sample variance) compared to true values of $-2.3$ and $-2.89$ respectively. Due to the sampler assuming different chemistry to that of the data, a run of \textit{Chempy} using the true values of the SSP and ISM parameters will not reproduce the observational abundances exactly, even in the absence of observational errors. Instead, it is likely that a closer match between \textit{Chempy} predictions and observations will be obtained using a slightly different set of parameters, leading to a bias in the derived posterior parameters. This is partially ameliorated by the inclusion of free model errors, which have the effect of downweighting elements that fit the data less well. If these are not implemented (i.e. setting $\sigma_\mathrm{model}^j=0$ for all $j$), the fractional bias is significantly increased, giving $\widehat{\alpha}_\mathrm{IMF}=-2.374\pm0.005$ and $\widehat{\log_{10}(N_\mathrm{Ia})}=-3.11\pm0.01$ for $n_\mathrm{stars}=200$, demonstrating their utility for real analyses. In addition, when the true yield set is not known, the bias may be approximated by rerunning the inference multiple times with different yield tables to give an empirical `yield set bias' that can be combined with the sources of uncertainty discussed above.

Fig.\,\ref{fig: model_errors}b and Tab.\,\ref{tab:model_errors}b show the posterior distributions of the $\{\sigma_\mathrm{model}^j\}$, as in the previous section. Unlike before, we observe a strong preference for non-zero model errors, especially for C, N and Si abundances, which have median values significantly greater than the observational errors ($0.05$\,dex). This indicates that our model is unable to reproduce the observed abundances of these elements. In all three cases, we have significant differences between the alternative and TNG yields in the dominant nucleosynthetic process (cf.\,Fig.\,\ref{fig: wrong_yield_comparison}), justifying these results.\footnote{\resub{We cannot directly identify the elements with the largest model errors to those with the largest differences in Fig.\,\ref{fig: wrong_yield_comparison} since \textit{Chempy} abundances are a function of the yields across all metallicities and times, whilst the figure shows the output of a single SSP at solar metallicity. In addition, the model errors are affected by the constraining power of individual elements on the SSP and ISM parameters; incorrectly produced elements that affect the posterior constraints more strongly will have larger model errors.}} In contrast, the model errors for [He/Fe] and [Fe/H] are small, indicating that there is little change to these abundances caused by changing yield set, again consistent with Fig.\,\ref{fig: wrong_yield_comparison} (also noting that, even at late times, most of the H and He comes from the primordial gas). From the table, we note that the fractional widths of the posterior distributions shrink as $n_\mathrm{stars}$ increases, whilst the median values increase for small $n_\mathrm{stars}$ then become independent of the sub-sample size. For small sub-samples, it is tempting to think that the model errors will be large since there will be stars whose abundances cannot be well reproduced by the model. However, in this limit, we have a large number of free parameters to constrain with very little data, so any such errors can easily be absorbed into an ISM or SSP parameter, and the distributions will tend to reproduce the priors. As the number of data-points becomes large, the data-set becomes far more constraining, and we can effectively distinguish between SSP, ISM and model error effects, causing the model error distributions to settle about their preferred values.

This analysis shows that to avoid bias in the inference of the galactic IMF and SN\,Ia parameters, we require yield sets that accurately represent galactic chemistry. Introduction of the model error parameters helps with this, as it allows the sampler to place greater weight on more well reproduced elements, reducing the bias to $\sim 3\%$ in this instance, despite significant differences between yield tables. Further assistance is provided by making informed choices about the yield tables, e.g. using those that best recover observational data-sets such as the proto-solar abundances \citepalias{2018ApJ...861...40P}, and restricting to elements that are known to be well-fit by current models \resub{\citep{2019ApJ...874..102W,2019arXiv190806113G}. In observational contexts, we would additionally exclude elements such as C and N which are known to undergo significant changes in their abundance during stellar evolution \citep{2000A&A...354..169G,2019A&A...621A..24L}.} A further benefit of the model errors is as a diagnostic tool; in analysis of observational data, we can assess how well individual yields match reality via the magnitude of $\sigma_\mathrm{model}^j$ and, in the (futuristic) case of highly accurate nucleosynthetic models, uncover observational biases.

\subsection{Mock Data from IllustrisTNG}\label{subsec: mocks_TNG}
The simplified ISM physics parametrization used in \textit{Chempy} does not accurately describe the physical Universe. To explore the biases in the inferred galactic parameters caused by this, we apply the analysis of Sec.\,\ref{sec: sampling} to mock data drawn from the vastly more complex IllustrisTNG simulation, which was described in Sec.\,\ref{sec: GCE_models}a.

Here, we extract a single galaxy from the $z = 0$ snapshot of the highest-resolution TNG100-1 simulation, choosing a subhalo (index 523071) with mass close to $10^{12}\,\mathrm{M}_\odot$, assuming this to be similar to the Milky Way (MW). From this, we extract 200 `stellar particles' from the $\sim 40,000$ present, each of which has mass $\sim 1.4\times 10^6 \mathrm{M}_\odot$ \citep{2019ComAC...6....2N}. These act as proxies for stellar environments, giving the elemental mass fractions, $\{d_i^j\}$, and cosmological scale factor, $a_i$, at the time of stellar birth. Mass fractions are converted to [X/Fe] abundance ratios using \citet{2009ARA&A..47..481A} solar abundances as in \textit{Chempy}, with the scale-factor ($a_i$) to birth-time ($T_i$) conversion performed using \texttt{astropy} \citep{astropy:2013,astropy:2018},\footnote{\href{http://www.astropy.org}{http://www.astropy.org}} assuming a $\Lambda$CDM cosmology with \citet{planck2015} parameters, as in TNG \citep{2018MNRAS.475..648P}.\footnote{As for the \textit{Chempy} mock data, we exclude any particles with $T_i\notin[2,12.8]\,\mathrm{Gyr}$ to ensure that the true times are well separated from our training age limits, avoiding neural network errors. This removes $\sim5\%$ of the stars.} Observational errors are incorporated as above, giving a full data-set that is identical in structure to the \textit{Chempy} mock data.

\begin{figure}
    \centering
    \includegraphics[width=0.5\textwidth]{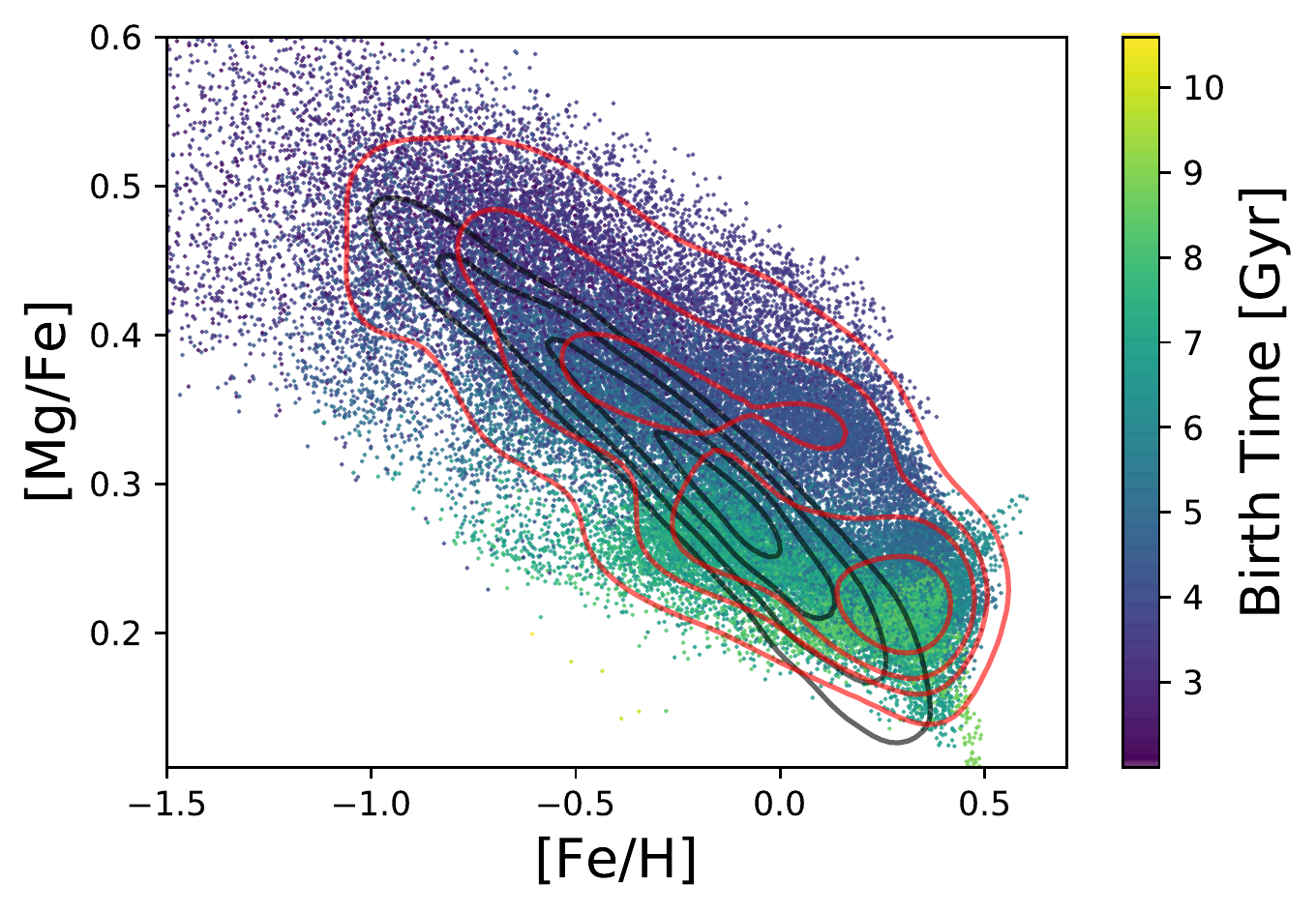}
    \caption{Chemical evolution tracks in the [Mg/Fe] vs. [Fe/H] plane for `stellar particles' taken from a Milky Way-like IllustrisTNG galaxy \citep{2018MNRAS.475..648P}, colored as a function of their birth-time $T_i$. This shows $\sim 40,000$ individual `stellar particles', with smoothed contours at $1$ to $4\sigma$ shown in red. For comparison, we plot smoothed contours of the \textit{Chempy} abundance distribution in black, using TNG yields and fixing the global parameters ($\alpha_\mathrm{IMF}$ and $\log_{10}(N_\mathrm{Ia})$) to the TNG values of $-2.3$ and $-2.89$ respectively, as in Sec.\,\ref{subsec: mocks_Chempy}. Contours are created from 1000 runs of \textit{Chempy}, drawing the local (ISM) parameters from the priors on $\vec\Theta_i$ (Tab.\,\ref{tab:priors}), and the birth-times, $T_i$, from the SFR model, assuming prior parameters (see Sec.\,\ref{subsec: chempy_intro}). We caution that these are \textit{prior} abundance predictions for \textit{Chempy} with no fitting performed, and that each TNG stellar particle contains a range of different mass (and lifetime) stars formed at the same time and composition. }
    \label{fig: tng_chempy_evolution_track}
\end{figure}

Fig.\,\ref{fig: tng_chempy_evolution_track} shows the chemical evolution tracks in the [Mg/Fe] vs. [Fe/H] plane for the full set of TNG stellar particles from the chosen galaxy. For comparison, we plot (black) contours obtained from a sample of 1000 \textit{Chempy} mock data-points (cf.\,Sec.\,\ref{subsec: mocks_Chempy}), with birth-times drawn from the range $[0,13.8]$\,Gyr, weighted by the \textit{Chempy} SFR prior, each with a random realization of the local parameters, $\vec\Theta_i$, sampled from the priors (Tab.\,\ref{tab:priors}).\footnote{Note that we do not convolve the SFR with the stellar lifetime function to create the \textit{Chempy} data for this plot. This is because we do not have individual stellar data for TNG, only the initial abundances and birth-times of large stellar particles, which contain many individual stars of varied lifetimes and masses.} The abundance distributions are broadly similar between the two simulations (as expected, since they utilize the same nucleosynthetic yields), though we note that the variance of the TNG data is much greater, especially along the [Fe/H] axis (analogous to the results of \citetalias[Fig.\,7]{2018ApJ...861...40P} which used a similar hydrodynamical simulation). Mismatches between the simulations are likely to result from the different ISM physics parametrizations, with TNG employing a far more realistic engine than the simple one-zone model of \textit{Chempy}. A major difference is in the SFR; this is set as a one-parameter $\Gamma$-distribution in \textit{Chempy}, but arises naturally from hydrodynamical processes in TNG. It is pertinent to note that the \textit{Chempy} ISM parameters used in Fig.\,\ref{fig: tng_chempy_evolution_track} are chosen without knowledge of the TNG simulation; better agreement can be found by using the posterior parameters for a data-set, though this is costly to do for a large number of stars.

\resub{The TNG galaxy used here was deliberately chosen to have both a high-$\alpha$ and low-$\alpha$ chemical evolution sequence (as observed in Fig.\,\ref{fig: tng_chempy_evolution_track}) to test our inference on a mock galaxy with MW-like properties. While recent simulations differ on the exact details of how bimodality develops, it is generally attributed to gas-rich mergers and different modes of star formation \citep{2018MNRAS.474.3629G,2018MNRAS.477.5072M,2019MNRAS.484.3476C,2019arXiv190909162B}. In chemo-dynamical models, bimodality similar to the MW can also be achieved by a combination of radial migration and selection effects without the need for mergers or starbursts \citep{2009MNRAS.396..203S,2013A&A...558A...9M,2017ApJ...835..224A}.
In the parametrization used here, \textit{Chempy} can assign each star to its own ISM environment, but cannot exchange gas between environments and has no sudden star formation or infall events. We hence investigate here whether this significantly biases our inference of the SSP parameters (noting that results from \citet{2019ApJ...874..102W} justify the treatment of ISM parameters as latent variables).}


The posterior distributions of $\vec\Lambda$ obtained from HMC sampling for the TNG data-set are shown in Fig.\,\ref{fig: mock_comparison}c and Tab.\,\ref{tab:results}c. As before, the sample and statistical variances are seen to decrease as $n_\mathrm{stars}$ increases, with the parameter estimates becoming statistical variance limited by $n_\mathrm{stars}\approx 10$. For $n_\mathrm{stars} = 1$, the statistical variance of the global parameters is similar to that found in the TNG studies of \citetalias{2018ApJ...861...40P}, which used the same chemical elements and yield tables, albeit with a different stellar data-set, leading to a different median $\vec\Lambda$ estimate. We note a generally larger sample variance for the TNG results compared to those in previous sections; this implies that the TNG mock data-set contains a broader range of stellar ISM environments than the \textit{Chempy} mock data, most likely because we are not limited by the simple \textit{Chempy} parametrizations. This is also demonstrated in Fig.\,\ref{fig: tng_chempy_evolution_track}, where the abundance-space distribution of TNG is seen to be much broader than that of the \textit{Chempy} priors. If a stellar particle outside the main \textit{Chempy} realm is included in the data-set by chance, the IMF slope is forced to shift to move the \textit{Chempy} abundance track, leading to a greater sample variance.

For all values of $n_\mathrm{stars}$ tested, there is good agreement between the inferred parameters and their true values, obtaining best estimates of $\widehat{\alpha}_\mathrm{IMF}=-2.283\pm0.007$ and $\widehat{\log_{10}(N_\mathrm{Ia})} = -2.889\pm0.008$ with 200 stars, highly consistent with TNG.\footnote{Note that this behavior is not simply the variables reproducing the priors; the $\log_{10}(N_\mathrm{Ia})$ prior was set as $-2.75\pm0.30$ which is very different to the above distribution.} In addition, the posterior estimates of $\vec\Lambda$ from individual sub-samples are consistent with the true values (to within $2\sigma$) for $n_\mathrm{stars}\gtrapprox 10$, though we caution that deviations exceeding $3\sigma$ are found when using only single stars in the analysis. For completeness, we display the full corner plot of the ten global parameters using $n_\mathrm{stars}=200$ in appendix \ref{appen: full_corner_plot}.

To place our results in an observational context, we additionally show the constraints on $\alpha_\mathrm{IMF}$ obtained from modern analyses using star counts in M31 \citep{2015ApJ...806..198W} and the Milky Way \citep{2019ApJ...870...44H}, as well as on $\log_{10}(N_\mathrm{Ia})$ from various observations of SN\,Ia \citep{2012MNRAS.426.3282M,2017ApJ...848...25M}. Whilst the centers of these constraints are clearly inconsistent with our results (since they use observational data, whilst we limit ourselves to a simulation), we may readily compare the widths of the contours to assess the constraining power of the various methods. Considering both sampling and statistical errors, our analysis gives stronger posterior constraints than the observational studies for both parameters, using $n_\mathrm{stars}\gtrsim 20$. Even when we account for modeling biases (e.g. in the case of incorrect yield tables), the technique of constraining galactic parameters from individual chemical element abundances is certainly competitive.


The model errors (Fig.\,\ref{fig: model_errors}c and Tab.\,\ref{tab:model_errors}c) exhibit similar trends with $n_\mathrm{stars}$ as discussed in previous sections. In this case however, we note small errors (below the observational error of $0.05\,$dex) for all abundances involving metal ratios, yet large errors ($\sim 0.2\,$dex) for [He/Fe] and [Fe/H] (becoming tightly constrained at large $n_\mathrm{stars}$). The former shows that the metal ratios are strongly constraining (especially [N/Fe] and [Si/Fe] in this case), but the latter indicates a mismatch between TNG and \textit{Chempy} either in terms of non-metal enrichment or the total metallicity (tracked by the ratio of metals to non-metals), which is consistent with the anomalous [Fe/H] behavior in Fig.\,\ref{fig: tng_chempy_evolution_track}. This discrepancy will be sourced by the difference in ISM physics between the simulations; whilst the metal ratios are set mainly by the chemical yields, the absolute metallicity depends strongly on details such as the stellar feedback strength and star formation history, which are difficult to encapsulate within \textit{Chempy}'s simple ISM physics parametrization. A likely cause of this difference is that we assume both AGB and SNe events to immediately deposit the same fraction of stellar feedback into the local ISM (i.e.\,$\mathrm{x}_\mathrm{out}$), which is unlikely due to the large differences in kinetic energy between the two processes. In TNG, the hotter SN feedback will be spread out far more and take more time to cool, whilst the colder AGB expulsions will be readily available to form new generations of stars. This will significantly affect the non-metal fractions in the simulation.
One way in which to ameliorate these problems would be by introducing additional free parameters into the \textit{Chempy} model, for example including separate AGB and SNe feedback fraction parameters or controlling the size of the simulation gas reservoir. Whilst this would likely reduce the model errors in [Fe/H] and [He/Fe], it would be at the expense of additional computation time, particularly if the parameters are chosen to be local, thus it has not been explored here. In our analysis, these issues are of limited importance, since the large size of [Fe/H] and [He/Fe] model errors diminishes the impact of these abundances in the likelihood analysis. Repeating the $n_\mathrm{stars}=200$ inference \textit{without} the model errors gives $\widehat{\alpha}_\mathrm{IMF}=-2.279\pm0.005$ and $\widehat{\log_{10}(N_\mathrm{Ia})} = -2.881\pm0.007$, showing a slight bias and $\sim4\sigma$ tension in the IMF parameter due to the poorly reproduced [Fe/H] and [He/Fe] abundances.

In terms of the local parameters, the posterior distributions show similar behavior to that of the \textit{Chempy} mock data (Sec.\,\ref{subsec: mocks_Chempy}). We observe a fractional error in the median inferred birth-times compared to their true values of $0.00^{+0.20}_{-0.24}$ ($0.01^{+0.19}_{-0.17}$) with a fractional posterior width of $0.17^{+0.01}_{-0.02}$ ($0.16^{+0.01}_{-0.02}$) for $n_\mathrm{stars}=1$ $(n_\mathrm{stars}=200$), only marginally narrower than the prior width of 20\%. Using only eight elements in the analysis, this technique is \textit{not} capable of providing precise estimates of stellar ages (or analogously local ISM parameters), yet it is clear that we can obtain strong constraints on the global parameters utilizing only weakly informative priors.

Considering the entirely different parametrizations of ISM physics between the two GCE models, our inferred SSP parameters are in impressive agreement with the true values. It is pertinent to note however, that the posterior confidence intervals on $\vec\Lambda$ are expected to shrink to zero as $n_\mathrm{stars}\rightarrow\infty$, as we do not include contribution to the variances from the errors made by \textit{Chempy}, thus we do expect a small bias to become apparent for very large $n_\mathrm{stars}$. Due to this, extension of the method to larger $n_\mathrm{stars}$ would be an interesting avenue of research. This is non-trivial however, since the sampling time becomes large (several hours on multiple cores) for $n_\mathrm{stars}\gtrsim 50$, thus we must look to alternative (approximate) sampling methods such as ADVI, allowing us to use many more data-points to ensure that error is dominated by systematics alone. 

\resub{\subsection{Potential Future Work}}
\resub{We briefly outline additional modifications that may need to be considered for our method to be applied to observational data. The largest obstacle arises from the uncertainties in the underlying nucleosynthetic yields, and advancement therein will improve the accuracy of the inference. This may take many forms, for instance with the usage of empirical yields \citep[e.g.][]{2012AcA....62..269A,2017MNRAS.467.1140J,2018MNRAS.474.4010B,2018MNRAS.475.1410P,2019arXiv190710606N}, the inclusion of the latest yield sets \citep[e.g.][]{2018MNRAS.476.3432P}, the implementation of binary star evolution effects \citep[e.g.][]{2015A&A...581A..22A,2017hsn..book..649B,2019A&A...626A.127J} or the propagation of nucleosynthetic yield uncertainties into our GCE model \citep{2016MNRAS.463.4153R}. Similarly, a more advanced error treatment will help to reduce bias from inevitably imperfect models. With some modification, our statistical analysis may itself be extended to infer empirical yields for nucleosynthetic processes, albeit with the loss of neural net functionality and therefore speed.}

\resub{Further improvements can be made by broadening the set of elements used, made possible by adding more nucleosynthetic channels, such as neutron-star mergers \citep{2017ApJ...836..230C} or sub-Chandrasekhar SNe\,Ia \citep{2011ApJ...734...38W,2018ApJ...854...52S}. These will also give tight constraints on the frequency of these additional channels. In observational contexts, we are limited to use only elements that do not undergo significant post-birth changes in abundance; inclusion of a model that maps the observed stellar elemental abundances to their birth abundances \citep[e.g.][]{2017ApJ...840...99D} would allow a greater number of elements to be used. Furthermore, increasing $n_\mathrm{stars}$ would allow us to add more free variables, for instance SN\,Ia time-delay parameters, process dependent outflow fractions, free solar abundances, and more complex (or hierarchical) star formation histories. The current precision of stellar age estimates does not seem to be a limiting factor for our method, especially since this is marginalized over, though more precise estimates would be expected to somewhat reduce the uncertainty on the SSP parameters.}

\resub{When choosing a set of stars to analyze, it is important to consider the selection function \citep[e.g.][]{2016A&A...589A..66H, 2016AN....337..880J}, and a study using only thin or thick disk stars may give us valuable insight into its effects. In our analysis of global SSP parameters however, it appears to be sufficient to cover a large variety of the abundance space without the need for exhaustive knowledge of the selection function. This is in agreement with the work of \citet{2019ApJ...874..102W}, which notes that a given star's abundances will carry the imprint of the global parameters and nucleosynthetic yields. Additional improvement may also be achieved by the use of Mg as the normalization element in the \textit{Chempy} likelihood rather than Fe, as in \citet{2019ApJ...874..102W}.}

\resub{Whilst this study has begun to explore the effects of modelling simplifications and incorrect yield tables, we caution that only a single set of analyses was run in each case, and is by no means intended as an exhaustive test to determine the applicability for the real MW. Other tailored tests will be necessary, for example performing a detailed analysis of how chemical evolution modeling assumptions can bias the results \citep{2017ApJ...835..128C}, or investigating the impacts of more complex sub-grid physics in the hydrodynamical model, such as a metallicity dependent IMF \citep{2019MNRAS.482..118G}.}


\section{Conclusions}\label{sec: conclusion}
In this paper, we have demonstrated a technique for inferring global galactic parameters controlling the SN\,Ia normalization, $\log_{10}(N_\mathrm{Ia})$, and the \citet{2003PASP..115..763C} IMF high-mass slope, $\alpha_\mathrm{IMF}$, using only stellar chemical abundance and age data. This builds upon previous work by the extension to multiple stars, which requires a more sophisticated statistical model and sampling technique. \resub{The inference technique is both fast and flexible, allowing strong constraints to be placed on global parameters using a large number of stars in a few tens of CPU-hours.}

Our core model has been the flexible `leaky-box' galactic chemical evolution (GCE) code \textit{Chempy} \citepalias{chempy}, used to predict elemental abundance ratios which are compared to observational data in a Bayesian framework. The \textit{Chempy} model requires input parameters describing both global and local physics, with the latter being specific to a star's \resub{formation environment}. Forming a statistical model for multiple stars has thus required each star to carry its own set of ISM parameters, all of which must be marginalized over. The star's birth-time was treated as an extra free parameter, which was also marginalized over given some initial estimate. In addition, we included a `model error' parameter for each chemical element, which can account for inaccuracies in \textit{Chempy}, for example from incorrect chemical yield tables. This allowed the sampler to dynamically downweight elements that fitted the data less well, reducing the bias in the global parameter estimates.

To allow for efficient sampling of the many-star posterior function, \textit{Chempy} was replaced by a neural network, trained to reproduce output chemical abundances given some initial parameter set \citepalias[cf.][]{2018ApJ...861...40P}. This converts \textit{Chempy} into a simple, and differentiable, analytic matrix function allowing us to use modern statistical methods to sample the high-dimensional posterior, in this case Hamiltonian Monte Carlo methods \citep{2012arXiv1206.1901N}. The full analysis pipeline has been made publicly available with a comprehensive tutorial \resub{\citep{oliver_philcox_jan_rybizki_2019}}.\footnote{\href{https://github.com/oliverphilcox/ChempyMulti}{github.com/oliverphilcox/ChempyMulti}}

Our analysis routine was tested using mock data; first with a data-set computed by \textit{Chempy} to test the neural network and sampling, augmented with broad observational errors of $5\%$ ($20\%$) in abundance (age). As the number of stellar data-points, $n_\mathrm{stars}$, increased, the estimated values of the SN\,Ia normalization and IMF slope were found to converge to the true values at high precision ($\lesssim1\%$ for individual data-sets with $n_\mathrm{stars}\gtrapprox50$). When using few stars, we observed significant sample variance in the derived parameter estimates between data-sets, indicating that caution must be used when interpreting inference results in single star analyses such as \citetalias{chempy}.

To explore the bias created by assuming incorrect chemical yields, we similarly analyzed a data-set created with a different set of yield tables, which was shown to give a bias of $\sim 3\%$ ($\sim 8\%$) in the posterior parameter estimates when model errors were (were not) included. This bias can be lowered by only using elements which are well predicted by our yield tables. \resub{Elements with larger model errors broadly corresponded to those with greater discrepancies between the yield tables}, showing the utility of model errors as a diagnostic tool \resub{for determining how well model yields represent the Universe's chemistry. In applications of this method to observational data, the analysis can be repeated with several different sets of yield tables to determine the bias empirically.}

Using a mock data-set drawn from a Milky-Way like galaxy in the IllustrisTNG \citep{2018MNRAS.473.4077P} simulation (which has known values of the global parameters and yields), we were able to test the bias in the parameter estimates caused by the ISM physics simplifications in \textit{Chempy}. \resub{These assumptions cause the outputs of \textit{Chempy} to span only a limited subset of abundance space; a point outside the typical \textit{Chempy} range may thus be expected to bias the inference results. In practice,} this was found to be insignificant, with posterior parameter estimates consistent with the true values across the range of data-set sizes tested. For $n_\mathrm{stars}=100$ we obtained constraints of $\alpha_\mathrm{IMF}=-2.283\pm0.010\,\mathrm{(statistical)}\pm 0.006\,\mathrm{(sample)}$ and $\log_{10}(N_\mathrm{Ia}) = -2.889\pm0.011\,\mathrm{(statistical)}\pm0.004\,\mathrm{(sample)}$ compared to true values of $-2.3$ and $-2.89$ respectively. This is highly competitive when compared to canonical galactic parameter studies such as star counts in M31, which give $\alpha_\mathrm{IMF} = -2.45^{+0.06}_{-0.03}$ \citep{2015ApJ...806..198W}.

The model errors showed the metal abundance ratios to be highly consistent between IllustrisTNG and \textit{Chempy}, yet there were large discrepancies for [Fe/H] and [He/Fe], suggesting that \textit{Chempy} is a relatively poor estimator of the overall metallicities (likely caused by our assumptions that AGB and SNe have the same feedback fraction to the local ISM and the feedback is accessible to new star formation immediately) though large model errors meant that these elements did not contribute significantly to the overall likelihood. \resub{We note that our inference was \textit{not} able to place strong constraints on stellar ages; this can be improved by using a greater number of elements in the analysis.}


The natural extension of this is the application to real data-sets, for example to red giant abundances from the APOGEE survey \citep{2016AN....337..863M}, combined with stellar age priors \citep[e.g.][]{2016ApJ...823..114N}. The statistical model remains the same in this context, yet we are subject to a number of sources of uncertainty, which, whilst partially ameliorated by our model error parameters, can bias our inference. As shown above, the choice of chemical elements and yield tables is of paramount importance, and one may make guided choices from studies such as \citet{2019ApJ...874..102W} and \citetalias{2018ApJ...861...40P} respectively. (Note also that we can obtain much stronger constraints on yield tables by using abundances from multiple stars, combining the techniques of \citetalias{2018ApJ...861...40P} with this work.) Furthermore, since we can only observe current stellar abundances, there can be biases due to post-birth changes in chemical abundances (significantly affecting elements such as C and N). \resub{Additionally, although the physics simplifications made by \textit{Chempy} were not found to have a large impact upon the TNG parameter constraints, this is not guaranteed for the real Universe. We are also sensitive to changes in the stellar lifetime functions and missing nucleosynthetic channels (e.g. neutron star mergers).}


These setbacks notwithstanding, it is clear that, in tandem with additional constraints such as star counts \citep[e.g.][]{2015ApJ...806..198W,2019ApJ...870...44H}, the methods in this paper could be used to obtain strong constraints on crucial galactic parameters such as the high-mass slope of the ISM and the number of SN\,Ia in the galaxy. Using approximate sampling methods such as ADVI, analysis with $n_\mathrm{stars}\sim 1000$ will become possible, allowing us to rigorously exploit the huge volumes of chemical abundance data available. This will enable many probes of galactic physics, for example testing the metallicity dependence of the IMF and attempting to infer the yield tables themselves.

\acknowledgments
We thank the following people for fruitful discussions; Robert Grand, Hans-Walter Rix, Rahul Dave, Chris Buswinka and Henry Wang. \resub{We are grateful to the anonymous referee for a detailed report which helped improve the clarity and impact of this work.} OHEP acknowledges funding from the Herchel-Smith foundation. JR acknowledges funding by the DLR (German space agency) via grant 50\,QG\,1403. 

%


\software{\resub{ChempyMulti \citep{oliver_philcox_jan_rybizki_2019}}, ChempyScoring \citep{oliver_philcox_2018_1247336}, Chempy \citep{2017ascl.soft02011R}, \texttt{scikit-learn} \citep{scikit-learn}, \texttt{astropy} \citep{astropy:2013,astropy:2018}, \texttt{PyMC3} \citep{pymc3}, \texttt{theano} \citep{theano}, \texttt{corner} \citep{corner} \& TikZ Bayesnet (\href{https://github.com/jluttine/tikz-bayesnet}{github.com/jluttine/tikz-bayesnet}).}

\bibliographystyle{aasjournal} 
\bibliography{jointlib} 



\appendix

\section{Neural Network Implementation}\label{appen: neural_network}
We here discuss the specifics of the neural network used in this analysis, which was introduced in Sec.\,\ref{sec: neural_nets}. The functional form is given by 
\begin{eqnarray}\label{eq: neural_net_def}
    \vec{h} &=& \vec{W}_0\cdot\vec{x} + \vec{b}_0 \\\nonumber
    \vec{y} &=& \vec{W}_1\cdot{}f(\vec{h}) + \vec{b}_1
\end{eqnarray}
for input vector $\vec{x}$ (dimension $n_\mathrm{in}$), output vector $\vec{y}$ (dimension $n_\mathrm{out}$),  and weights $\{\vec{W}_i$, $\vec{b}_i\}$, which are set via an optimizer during the network training. $\vec{h}$ represents the `hidden layer': a length $n_\mathrm{neuron}$ vector which is transformed by some vector-valued `activation function' $f$ before the output is constructed, allowing for the model to represent non-linear functions. It is here chosen as a $\tanh$ function.

\begin{figure}
\centering
\begin{minipage}[t]{.48\textwidth}
\vspace{0pt}
\centering
\begin{tikzpicture}

    \node[latent](x1){$x_1$};
    \node[latent,right=0.5 of x1](x2){$x_2$};
    
    \node[obs,below=1 of x2](h4){$h_{2,1}$};
    \node[obs,right=0.5 of h4](h5){$h_{2,2}$};
    \node[obs,right=0.5 of h5](h6){$h_{2,3}$};
    \node[obs,below=1 of x1](h3){$h_{1,3}$};
    \node[obs,left=0.5 of h3](h2){$h_{1,2}$};
    \node[obs,left=0.5 of h2](h1){$h_{1,1}$};
    
    \node[latent,below=1 of h2](y1){$y_1$};
    \node[latent,below=1 of h5](y2){$y_2$};
    
    \edge[->] {x1} {h1,h2,h3,h4,h5,h6};
    \edge[->] {x2} {h1,h2,h3,h4,h5,h6};
    
    \edge[->] {h1,h2,h3} {y1};
    \edge[->] {h4,h5,h6} {y2};
  
\end{tikzpicture}
\caption{Cartoon indicating the sparse neural network structure used in this analysis. We show a mock network with $n_\mathrm{in} = 2$ input nodes $\{x_i\}$ (representing \textit{Chempy} parameters) and $n_\mathrm{out} = 2$ output nodes $\{y_i\}$ (representing element abundances). Although there appear to be six hidden layer nodes (shown in gray), the $j$-th output node is connected to only $n_\mathrm{neuron} = 3$ hidden-layer nodes (labelled $h_{j,1}$, $h_{j,2}$, $h_{j,3}$), thus this structure is identical to a set of $n_\mathrm{out}$ fully-connected networks with only a single output node and $n_\mathrm{neuron}=3$. In the full analysis, we use $n_\mathrm{in} = 7$ (including a $T^2_i$ term), $n_\mathrm{neuron} = 40$ and $n_\mathrm{out}=8$, embedded in a similarly sparse structure, \resub{which was found to give better accuracy than a single fully-connected network}. Cartoon created using TikZ Bayesnet.}
\label{fig: network_cartoon}
\end{minipage}%
\hfill
\begin{minipage}[t]{.48\textwidth}
\vspace{0pt}
  \centering
  \includegraphics[width=\textwidth]{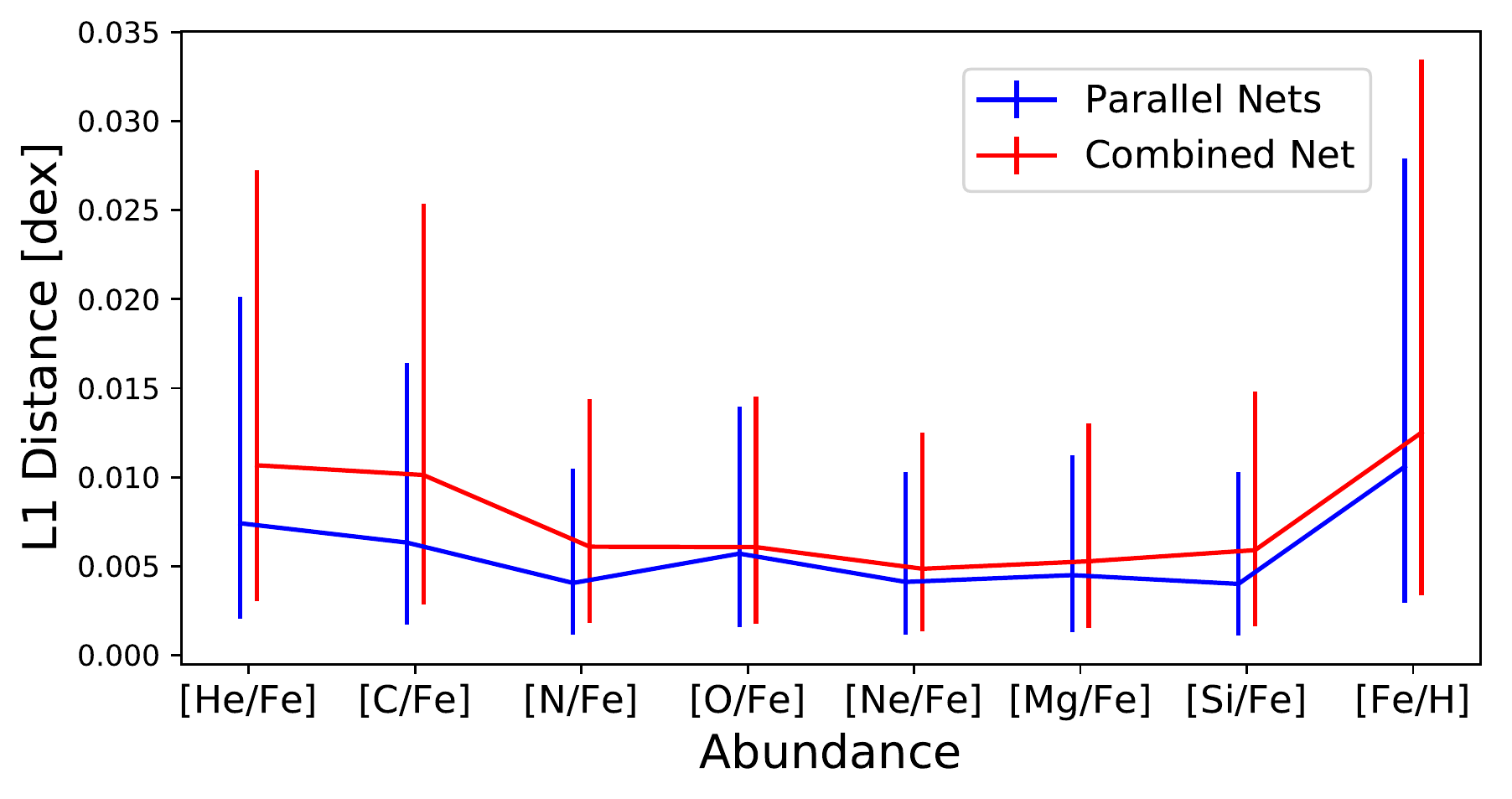}
    \caption{Absolute deviation between neural network predictions and true \textit{Chempy} abundances for 8 elements, computing distances from $5\times10^4$ parameter space samples, with inputs drawn from Gaussians centered at the \textit{Chempy} priors (Tab.\,\ref{tab:priors}) with widths of $2\sigma_\mathrm{prior}$. We show the median and \nth{16} / \nth{84} percentile deviations for two network configurations; using a single network for each element (blue) and using a joint network for all elements (red). Both instances are trained with $10^6$ data-points using $n_\mathrm{neuron} = 40$, and the former gives superior results.}
    \label{fig:network_element_error}
\end{minipage}
\end{figure}

There are a total of six inputs to the \textit{Chempy} function, from the global, local and birth-time parameters, as stated in Tab.\,\ref{tab:priors}. To allow for more accurate network fitting, we augment the input parameter vector with the value of $T_i^2$ (giving $n_\mathrm{in}=7$), which is useful since \textit{Chempy} has most complex dependence on $T_i$. Instead of creating a single large network with $n_\mathrm{out}=n_\mathrm{el}$ outputs, we here construct $n_\mathrm{el}$ individual networks with $n_\mathrm{out}=1$, allowing each element to be fit independently, giving greater network flexibility at smaller $n_\mathrm{neuron}$. This requires little additional computation time since the networks can be trained in parallel, and initial testing showed $n_\mathrm{neuron}=40$ to give sufficient network accuracy without overfitting. For later efficiency, the $n_\mathrm{el}$ fully-connected networks are combined into a single sparsely connected network (with a total of $n_\mathrm{el}n_\mathrm{neuron}$ hidden layer nodes), as illustrated in Fig.\,\ref{fig: network_cartoon}.

To teach the network to emulate \textit{Chempy}, we require a large volume of \textit{training data}; sets of input parameter vectors and associated output \textit{Chempy} abundances. Although a single run of \textit{Chempy} at a given output time $T_i$ already computes elemental abundances at 28 equally spaced time-steps, it is not pertinent to use these as 28 individual training points, since the resolution is low for the first few time-steps. Instead, we compute the model in full for each value of $T_i$ and take the final elemental abundances as training data, using a time-step of $T_i/28$. The training data-set is created from $1\times10^6$ random points in the six-dimensional parameter space (of $\vec\Lambda$, $\vec\Theta_i$ and $T_i$), with the SSP and ISM parameters being drawn from Gaussians (truncated for $\log_{10}(\text{SFR}_\mathrm{peak})$ as in Sec.\,\ref{subsec: chempy_intro}) centered at the prior-mean with $2\sigma_\mathrm{prior}$ width (cf.\,Tab.\,\ref{tab:priors}).\footnote{In \citetalias{2018ApJ...861...40P}, we created training data via a regular grid in parameter space. The new approach was found to give a faster converging network, and thus adopted here.} $T_i$ is drawn from a uniform distribution in $[1,13.8]\,$Gyr, ensuring good coverage over the relevant parameter space. This is the most computationally intense part of the analysis, with such a training set taking $\sim200$ CPU-hours to compute on a modern desktop machine, but can be trivially parallelized. For improved fitting, all network inputs and outputs are \textit{standardized}, with the new values $\hat{p}_i$ being derived from their unstandardized forms $p_i$ via
\begin{equation}\label{eq: standardization}
    \hat{p}_i = \frac{p_i-\mu_i}{\sigma_i},
\end{equation}
where $\mu_i$ and $\sigma_i$ are the mean and standard deviation of $p_i$. The uniformly distributed $T_i$ is instead linearly mapped to the interval $[0,1]$. This gives a total of $n_\mathrm{neuron}(n_\mathrm{in}+n_\mathrm{out}+1)+n_\mathrm{out} = 361$ free weight parameters for each of the $n_\mathrm{el}$ networks which are found by training with an `Adam' optimizer \citep{2014arXiv1412.6980K}, using a mean-square-error (L2) loss function and an adaptive learning rate, reducing as the training loss plateaus. This was implemented using the \texttt{scikit-learn} package \citep{scikit-learn} in Python, with training taking $\sim1$ CPU-hours (but may be parallelized $n_\mathrm{el}$-fold).

\begin{figure*}
    \centering
    \includegraphics[width=0.99\textwidth]{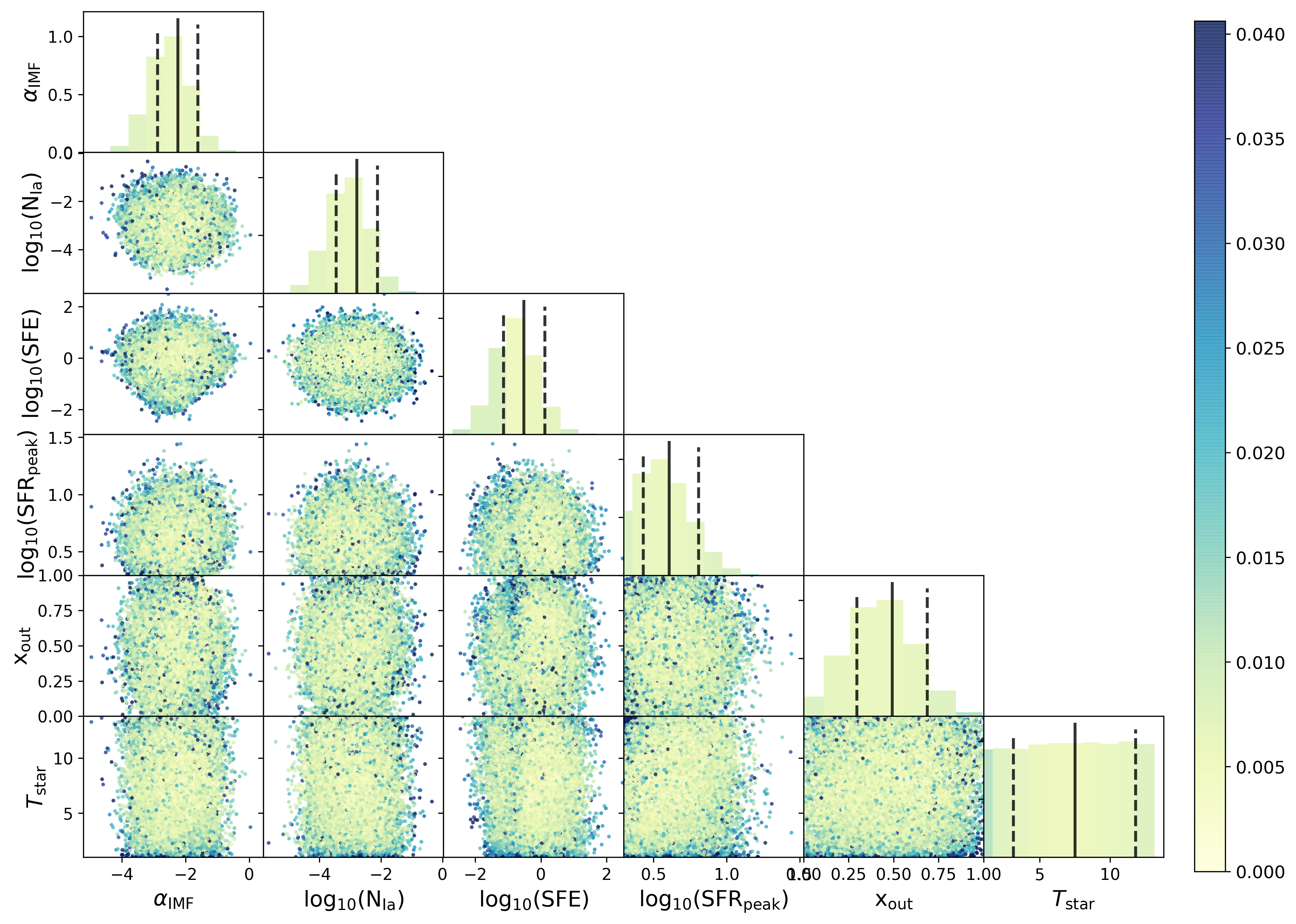}
    \caption{Mean neural network error across all elements as a function of position in the six-dimensional \textit{Chempy} parameter-space. The histograms on the diagonal show the distribution of test data-points, with their colors indicating the mean error in each bin. Full (dashed) lines indicate the median ($\sigma_\mathrm{train} = 2\sigma_\mathrm{prior}$) training values used in this sample. Off-diagonal plots show the marginal distribution of the error with respect to pairs of parameters. (Note that the $x_\mathrm{out}$ parameter is restricted to $[0,1]$, as in the full analysis, since values outside this region are unphysical.) \resub{The network errors are small in the center of parameter space (where the priors are concentrated) giving minimal bias to the inference.}}
    \label{fig:network_parameter_error}
\end{figure*}

Testing is performed by comparing the true abundances to the neural network predictions across an independent `test set' of $5\times10^4$ \resub{points (each consisting of an input parameter vector and a set of output abundances)}, computed as for the training data. Using the L1 distance metric (the absolute deviation between two values) we obtain a median error of $0.005_{-0.004}^{+0.008}$\,dex across the entire testing parameter space and $n_\mathrm{el}=8$ elements, well below typical observational errors of $0.05\,$dex, thus we take the network to be a good approximator of the \textit{Chempy} function. Figs.\,\ref{fig:network_element_error}\,\&\,\ref{fig:network_parameter_error} show the error as a function of the element and position in parameter space respectively, with the former also demonstrating the benefits from using individual networks for each element rather than a single fully-connected network. As expected, the network errors are small in the center of the distribution, but grow towards the edges of parameter space, where the function is sampled less finely. \resub{In particular, errors are greatest at the extremes of $T_\mathrm{star}$; for this reason we exclude stars with $T_\mathrm{star}\notin[1,13.8]\,\mathrm{Gyr}$ from the analysis, avoiding the need for a greater volume of training data.} If we required a more accurate network, this could be obtained using a large training data-set (possibly encompassing a greater prior width) or more neurons.

\section{Introduction to Hamiltonian Monte Carlo (HMC)}\label{appen: HMC_details}
We here present a broad overview of the HMC algorithm, which allows us to sample relatively high-dimensional posteriors with much greater efficiency than standard MCMC methods. In this paper, HMC is implemented via the \texttt{PyMC3} package \citep{pymc3}.

Following the notation of \citet{2013arXiv1312.0906B}, consider a posterior distribution $\pi(q)$ with parameter $q$, from which require samples. Instead of sampling $\pi(q)$ directly, we here introduce a `momentum' parameter $p$ and sample the joint density $\pi(p,q)=\pi(p|q)\pi(q)$, for user-defined conditional distribution $\pi(p|q)$ (often chosen as a Gaussian). In line with classical mechanics, we introduce a Hamiltonian density 
\begin{equation}
    H(p,q) = -\log\pi(p,q) = T(p|q)+V(q),
\end{equation}
identifying the kinetic and potential energies $T(p|q) = -\log\pi(p|q)$ and $V(q) = -\log\pi(q)$ respectively. (The kinetic energy becomes a simple quadratic in $p$ if we choose a Gaussian for $\pi(p|q)$.) 

Given this identification, we sample a value of the momentum $p$ from the conditional distribution $\pi(p|q)$ then evolve the variables $p$ and $q$ for some period of time according to Hamilton's equations for $H(p,q)$;
\begin{equation}
    \frac{\mathrm{d}q}{\mathrm{d}t}  = \frac{\partial H}{\partial p}, \quad \frac{\mathrm{d}p}{\mathrm{d}t} = -\frac{\partial H}{\partial q},
\end{equation}
requiring solution of a first-order differential equation (usually via leapfrog methods). After some number of time-steps, a new value of $p$ is drawn and the process repeated, with the individual samples of $q$ at each time-step forming the posterior chain. This results in a much more efficient sampling of the parameter space than just making random jumps in $q$ (as in conventional MCMC algorithms), since we additionally use the gradients of $H$ with respect to $p$ and $q$. Notably, this requires differentiability of the posterior $\pi(q)$, which limits the utility of HMC in many astrophysical contexts.

One pitfall of HMC is the addition of multiple free-parameters controlling the number and size of integration steps that should be taken from a given starting $(p,q)$ before a new momentum $p$ is drawn, which could require difficult tuning. This is solved with the No U-Turn Sampler \citep[NUTS; ][]{2011arXiv1111.4246H}, which \resub{(a) provides a physically motivated way in which to compute the step-size and (b) finds the optimal number of integration steps by integrating Hamilton's equations both forwards and backwards in time until the path in phase-space doubles-back on itself (and hence stops producing useful samples)}. Although HMC provides a large reduction in computation time compared with standard MCMC approaches, we can still encounter difficulties for very complex or high-dimensional posteriors, with the sampler taking too long to converge. For the analysis presented above, restricting to sampling times less than a few hours limits us to $n_\mathrm{stars}\leq 200$, though we are still able to produce high precision parameter estimates with this size of data-set. 

For more efficient sampling with large $n_\mathrm{stars}$ it may be more appropriate to use a HMC-within-Gibbs sampling approach, with HMC used to perform the parameter updates for $\vec\Lambda$, $\Sigma$ and $\{\vec\Theta_i,T_i\}$ separately, (as suggested in \citealt{2012arXiv1206.1901N}) although this has not been implemented here. As mentioned above, an additional possibility is to use approximate sampling methods such as `Automatic Differentiation Variational Inference' \citep[ADVI; ][]{2013arXiv1312.6114K,2016arXiv160300788K,2017arXiv170309194R}, which approximates the (possibly transformed) posterior function as a product of univariate Gaussians that can be trivially sampled from. This approximation depends on a number of latent parameters (describing the shape and location of each Gaussian), which are optimized via gradient-descent, again requiring differentiability. Whilst the assumption of Gaussianity may seem to be highly restrictive, it is often found to work well in practice, especially when we additionally allow for correlations between some or all parameters (in `Full Rank' ADVI, in contrast to the standard `Mean Field' ADVI). Whilst not considered in this paper, this may be useful for analyses containing a greater number of model parameters, for instance if the chemical yields are also left free.

\section{Full Global Parameter Corner Plot}\label{appen: full_corner_plot}
\begin{figure}
    \centering
    \includegraphics[width=0.99\textwidth]{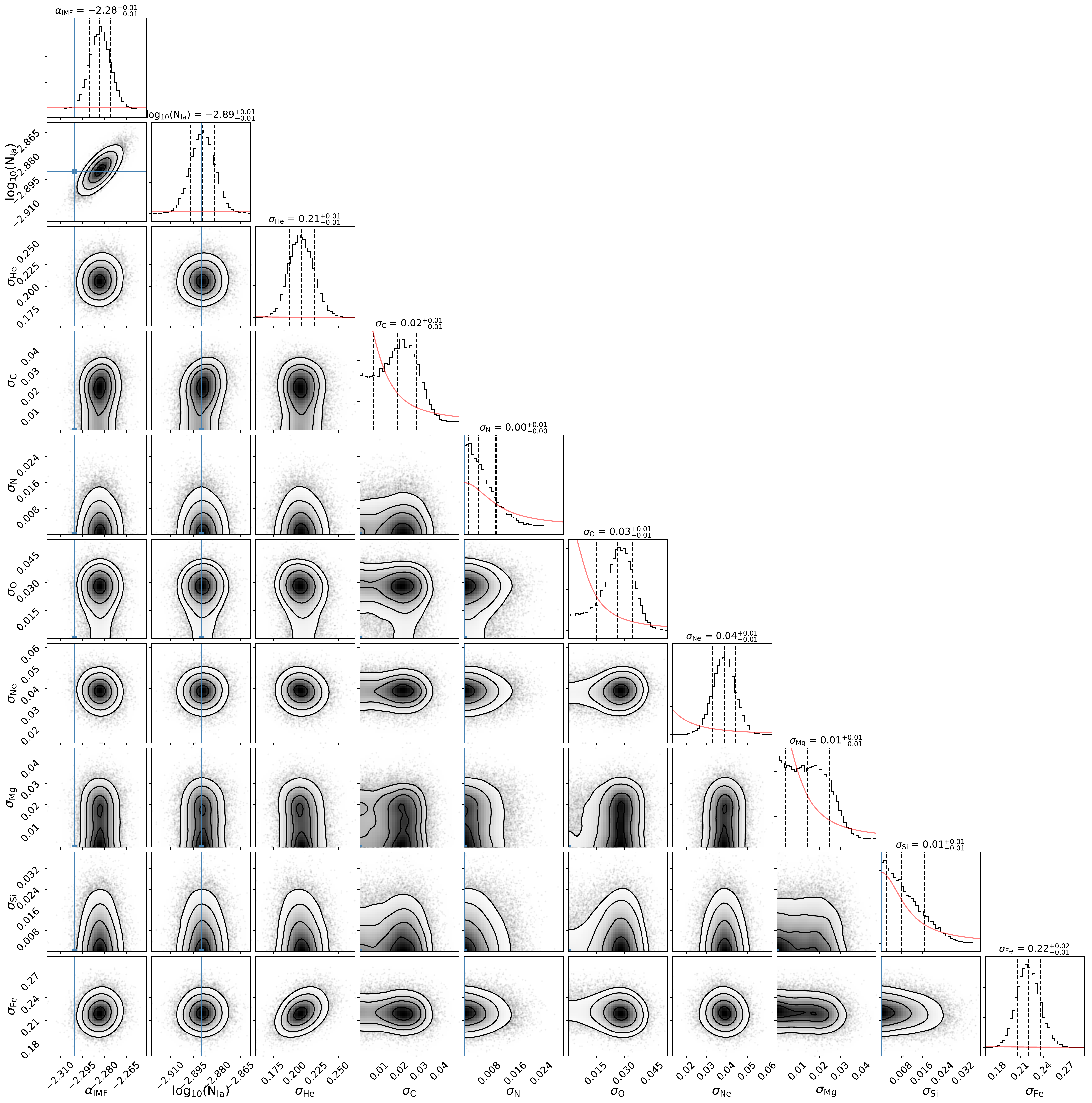}
    \caption{Corner plot illustrating part of the sampled posterior function using $n_\mathrm{stars}=200$ mock IllustrisTNG mock data-points, from $1.6\times10^4$ posterior samples obtained using HMC methods. We display only portions corresponding to the global SSP parameters, $\vec\Lambda=\{\alpha_\mathrm{IMF},\log_{10}(N_\mathrm{Ia})\}$, and model errors for each element $\Sigma=\{\sigma_\mathrm{model}^j\}$. The true values of $\vec\Lambda$ are marked in blue and are highly consistent with the SN\,Ia parameter, with a slight offset observed for $\alpha_\mathrm{IMF}$. Dashed lines in the one-dimensional histograms indicate the \nth{16}, \nth{50} and \nth{84} percentiles and smoothed contours (at 1 to 4$\sigma$ levels) are shown in the two-dimensional histograms. \resub{The prior distributions are indicated by red curves in the histograms.} Plot created using \texttt{corner} \citep{corner}.} 
    \label{fig: TNG_corner_full}
\end{figure}

Fig.\,\ref{fig: TNG_corner_full} shows the corner plot of the \textit{Chempy} posterior for HMC sampling of the TNG data-set using $n_\mathrm{stars}=200$, as discussed in Sec.\,\ref{subsec: mocks_TNG}. Since the full posterior exists in a 810-dimensional space, we show only the portions corresponding to the SSP parameters, $\vec\Lambda$, and model errors, $\Sigma = \{\sigma_\mathrm{model}^j\}$. Whilst the $\log_{10}(N_\mathrm{Ia})$ parameter is highly consistent with the true value, there is a slight tension in the $\alpha_\mathrm{IMF}$ parameter, though this may be caused by sample bias. The large non-zero errors of [Fe/H] and [He/Fe] (here denoted by $\sigma_\mathrm{Fe}$ and $\sigma_\mathrm{He}$) are clearly apparent, with the model error histograms matching those of Fig.\,\ref{fig: model_errors} and often close to the prior Half-Cauchy distributions. Furthermore, we note strong correlations between $\alpha_\mathrm{IMF}$ and $\log_{10}(N_\mathrm{Ia})$ (matching that found in \citetalias{chempy}), with a larger $\alpha_\mathrm{IMF}$ leading to more SN\,II, which require more SN\,Ia to obtain the correct abundance ratios of $\alpha$ and iron-peak elements. The model errors appear to be largely uncorrelated both with each other and with the SSP parameters, though there is weak correlation between $\sigma_\mathrm{Fe}$ and $\sigma_\mathrm{He}$ since both trace the overall metallicity of the simulation.


\end{document}